\documentclass[a4paper,fleqn,usenatbib]{mnras}

\usepackage{newtxtext,newtxmath}
\usepackage[T1]{fontenc}
\usepackage{ae,aecompl}

\usepackage{graphicx}	% Including figure files
\usepackage{amsmath}	% Advanced maths commands
\usepackage{amssymb}	% Extra maths symbols

\title[The MW's Disk of Classical Satellites in Gaia DR2]{The Milky Way's Disk of Classical Satellite Galaxies\\ in Light of Gaia DR2}

\author[M. S. Pawlowski \& P. Kroupa]{
Marcel S. Pawlowski,$^{1}$\thanks{Schwarzschild Fellow; e-mail: mpawlowski@aip.de}
and Pavel Kroupa$^{2,3}$
\\
$^{1}$Leibniz-Institut f\"ur Astrophysik Potsdam (AIP), An der Sternwarte 16, D-14482 Potsdam, Germany\\
$^{2}$Helmholtz-Institut f\"ur Strahlen- und Kernphysik, University of Bonn, Nussallee 14-16, D- 53115 Bonn, Germany\\
$^{3}$Charles University in Prague, Faculty of Mathematics and Physics, Astronomical Institute, V Hole\v{s}ovi\v{c}k\'{a}ch 2, CZ-180 00 Praha 8, \\Czech Republic
}

\date{Accepted 2019 November 7. Received 2019 November 7; in original form 2019 August 26}

\pubyear{2019}

\begin{document}
\label{firstpage}
\pagerange{\pageref{firstpage}--\pageref{lastpage}}
\maketitle

\begin{abstract}
We study the correlation of orbital poles of the 11 classical satellite galaxies of the Milky Way, comparing results from previous proper motions with the independent data by Gaia DR2. Previous results on the degree of correlation and its significance are confirmed by the new data. A majority of the satellites co-orbit along the Vast Polar Structure, the plane (or disk) of satellite galaxies defined by their positions. The orbital planes of eight satellites align to $<20^\circ$\ with a common direction, seven even orbit in the same sense. Most also share similar specific angular momenta, though their wide distribution on the sky does not support a recent group infall or satellites-of-satellites origin. The orbital pole concentration has continuously increased as more precise proper motions were measured, as expected if the underlying distribution shows true correlation that is washed out by observational uncertainties. The orbital poles of the up to seven most correlated satellites are in fact almost as concentrated as expected for the best-possible orbital alignment achievable given the satellite positions. 
Combining the best-available proper motions substantially increases the tension with $\Lambda$CDM cosmological expectations: $<0.1$\ per cent of simulated satellite systems in IllustrisTNG contain seven orbital poles as closely aligned as observed. Simulated systems that simultaneously reproduce the concentration of orbital poles and the flattening of the satellite distribution have a frequency of $<0.1$\ per cent for any number of $k > 3$\ combined orbital poles, indicating that these results are not affected by a look-elsewhere effect. This compounds the Planes of Satellite Galaxies Problem.
\end{abstract}


\section{Introduction}

Two independent studies, \citet{1976MNRAS.174..695L} and \citet{1976RGOB..182..241K}, first described a notable feature in the spatial distribution of the then-known satellite galaxies and distant globular clusters around the Milky Way: they align close to a common great circle on the sky, which is also traced by the Magellanic Stream. In three dimensions, these Milky Way satellites are found close to a common plane. This plane of satellites has since been termed the Vast Polar Structure (VPOS) of the Milky Way. It consists not only of satellite galaxies, but also globular clusters and streams were found to align \citep{2012MNRAS.423.1109P}. The VPOS is oriented nearly perpendicular to the Galactic disk and its spatial extent encompasses the whole virial volume of the Galaxy. The alignment of the Magellanic Stream -- as well as several other streams -- with the VPOS indicates that at least some of the satellites orbit along the common plane. This has been supported by early proper motion studies which indicated that several of the 11 classical satellite galaxies have orbital planes consistent with alignment with the VPOS, and are preferentially orbiting in the same sense \citep{2009MNRAS.394.2223M}. The satellite galaxies in the VPOS thus appear to not only be spatially but also kinematically correlated.

\citet{2005A&A...431..517K} were the first to point out the inconsistency between the highly flattened spatial distribution of the then-known 11 classical satellite galaxies of the Milky Way and the expected distribution of cosmological sub-structures. This has initiated a debate \citep{2005ApJ...629..219Z,2005MNRAS.363..146L,2007A&A...463..427P,2009MNRAS.399..550L,2010A&A...523A..32K, 2011MNRAS.415.2607D} about the degree of tension between the observed system and cosmological expectations based on the standard cosmological cold dark matter model ($\Lambda$CDM). 

It triggered numerous follow-up studies as additional satellite galaxies of the Milky Way were discovered, which largely confirmed the alignment already seen in the 11 classical satellites \citep{2007MNRAS.374.1125M,2009MNRAS.394.2223M,2014ApJ...790...74P,2015MNRAS.453.1047P}. This also motivated searches for similar structures around more distant host galaxies \citep{2006AJ....131.1405K, 2013AJ....146..126C,2014ApJ...788..146S,2014Natur.511..563I,2015MNRAS.449.2576C,2015ApJ...805...67I,2015MNRAS.453.3839P,2017A&A...602A.119M}, which culminated in the discovery of a very narrow plane consisting of about half of M31 satellite galaxies with coherent line-of-sight velocity trend indicative of a rotating plane \citep{2013ApJ...766..120C,2013Natur.493...62I}, and the existence of one or two satellite planes around the Centaurus\,A galaxy which also show an unexpected degree of kinematic coherence that is consistent with a rotating satellite plane \citep{2015ApJ...802L..25T,2016A&A...595A.119M,2018Sci...359..534M, 2019arXiv190702012M}.

Satellite systems as narrow as the observed ones and simultaneously as kinematically correlated are regularly reported to be exceedingly infrequent in cosmological simulations \citep{2014ApJ...784L...6I,2014MNRAS.442.2362P,2015MNRAS.452.3838C,2018MNRAS.478.5533F,2019MNRAS.tmp.1692S}. Numerous suggested solutions of this Planes of Satellite Galaxies Problem have been investigated, but none are generally agreed upon to satisfactorily address the issue. This includes the infall of satellite galaxies in groups \citep{2008MNRAS.385.1365L,2008ApJ...686L..61D,2009ApJ...697..269M,2013MNRAS.429.1502W,2018MNRAS.476.1796S}, the accretion of satellites along the cosmic web \citep{2011MNRAS.413.3013L,2011MNRAS.411.1525L,2012MNRAS.424...80P,2014MNRAS.442.2362P}, baryonic effects in cosmological simulations \citep{2015ApJ...815...19P,2017MNRAS.466.3119A,2018Sci...359..534M,2019ApJ...875..105P}, and special environments or properties of the host halos such as Local-Group-like pair configuration, halo concentration or formation time \citep{2014ApJ...789L..24P,2015MNRAS.452.1052L,2015ApJ...809...49B,2019ApJ...875..105P}. This makes the Planes of Satellite Galaxies problem one of the most long-standing small-scale problems of $\Lambda$CDM. The only presently known, viable physical process capable of forming highly phase-space correlated systems of dwarf galaxies and star clusters, including counter-orbiting satellites, is their formation in gas-rich tidal tails of interacting galaxies \citep{2011A&A...532A.118P,2012MNRAS.427.1769F,2014MNRAS.442.2419Y}. However, such tidal dwarf galaxies are dark-matter free in a $\Lambda$CDM universe \citep{2018MNRAS.474..580P,2019A&A...626A..47H}. This is at odds with the high mass-to-light ratios of many of the observed dwarf satellite galaxies, unless the dark matter hypothesis is abandoned in favor of a modified gravity approach \citep{2018A&A...614A..59B,2018MNRAS.477.4768B}. For a review of the observed satellite planes, comparison of them with cosmological simulations, and proposed solutions to the planes of satellite galaxies problem, see \citet{2018MPLA...3330004P}.

% These aren't the comments you're looking for.

In this debate, the Milky Way satellite system thus far is a special case since only for it full three-dimensional velocities can be obtained by combining spectrosopic measurements of their line-of-sight velocities with proper motion measurments giving the tangential velocity components. This allows one to test whether the orbits of the satellite galaxies are aligned with the plane defined by their current spatial positions, as expected for a dynamically stabilized structure. Indeed, the coherent kinematics are one of the main reasons for the strong tension between the observed satellite system and satellite systems in cosmological simulations which typically show a much lower degree of kinematic coherence \citep{2008ApJ...680..287M,2009MNRAS.399..550L,2014MNRAS.442.2362P,2014ApJ...789L..24P,2019MNRAS.tmp.1692S}.

Proper motion measurements have become increasingly accurate in the past decade, culminating in the revolutionary data provided by the second data release of the Gaia mission (Gaia DR2, \citealt{2016A&A...595A...1G, 2018A&A...616A...1G}). This rich data set enabled the measurement of the most precise proper motions to date for some of the classical Milky Way satellites \citep{2018AA...616A..12G, 2018A&A...619A.103F}, as well as the first proper motion measurements for many of the fainter satellites, bringing the total census of satellite galaxies for which some proper motion constrains are available to over 40 \citep{2018ApJ...863...89S, 2018A&A...619A.103F,  2018ApJ...867...19K, 2018A&A...620A.155M,  2019ApJ...875...77P}. 

In the context of the debate about the VPOS and the degree of tension between the orbital coherence of the 11 classical satellite galaxies and expectations derived from $\Lambda$CDM simulations, the Gaia proper motions provide an entirely independent data source to investigate the orbital coherence. Finding significantly different orbital coherence might indicate that random chance or unknown systematics have played a role in the previously reported kinematic coherence, and would thus imply that the tension with $\Lambda$CDM was overestimated.

It is possible to read the results and qualitative discussion of \citet{2018AA...616A..12G} in this way. They report that, now that Gaia DR2 has pinned-down the proper motions for the satellites, a relatively coherent phase-space distribution has been uncovered which, however, is not consistent with a single disk of satellites because not all of the classical satellite galaxies share a single plane of motion. It is understandable that these conclusions are sometimes interpreted as a blow to the observational evidence of a VPOS and thereby as exonerating the $\Lambda$CDM model of cosmology, despite the absence of a quantitative analysis supporting this interpretation.

% I got a bad feeling about this.

However, comparing the reported results of \citet{2018AA...616A..12G} with previous studies more carefully reveals that not having all of the 11 classical satellite galaxies participate in the coherent motion is in fact perfectly consistent with previous findings based on other proper motion measurements. For example, \citet{2013MNRAS.435.2116P} note a coherent alignment of 7 to 9 out of the 11 classical satellites, that the satellites {\it preferentially} co-orbit, and that the orbital poles of at least 6 of the 11 satellites are closely aligned. All these statements remain true in light of the Gaia DR2 proper motions as presented in \citet{2018AA...616A..12G}. A more detailed analysis of the implications of the spatial alignment of the observed 11 classical satellite galaxies as well as the satellite galaxies discovered by the Sloan Digital Sky Survey was performed in \citet{2016MNRAS.456..448P}. They compared the observed satellite positions with model satellite distributions consisting of a planar and an isotropic sub-population, while taking into account survey footprint biases. The study found that having more than 50\,per\,cent of all considered satellites as part of an isotropic distribution is excluded with 95 per cent confidence. However, this also implies that up to half of the considered satellite can have uncorrelated orbital poles, which would also be in line with the finding that only about half of M31's satellite galaxies are part of a planar structure \citep{2013Natur.493...62I}.

It is thus not immediately clear whether the Gaia DR2 proper motions confirms the previous results, strengthens the tension with $\Lambda$CDM, or alleviates the planes of satellite galaxies problem. To address this quantitatively, here we revisit the alignment of orbital poles of the 11 classical satellite galaxies of the Milky Way in light of the most recent proper motion measurements including the Gaia DR2, Hubble Space Telescope, and ground-based observations (Sect.  \ref{sect:orbitalpoles}). In particular, we investigate how the measured orbital coherence has evolved in light of increasingly precise proper motion measurements (Sect. \ref{subsec:timeevolution}), discuss the significance of the alignment of orbital poles and compare it to the best-possible alignment expected from the satellite positions alone (Sect. \ref{sec:Discussion}). We also address what effect the improved data has on the degree of tension between the $\Lambda$CDM model and the observed plane of satellite galaxies (Sect. \ref{sec:illustris}).

\section{Updated Orbital Poles for the 11 Classical Satellites}
\label{sect:orbitalpoles}

Following \citet{2013MNRAS.435.2116P}, we focus on the 11 classical satellite galaxies of the Milky Way. This provides a largely complete sample of the brightest satellite galaxies with $M_\mathrm{V} \leq -8.8$, which has frequently been used to compare to cosmological simulations \citep{2014MNRAS.442.2362P, 2014ApJ...789L..24P, 2019ApJ...875..105P,2013MNRAS.429.1502W,2019arXiv190402719S, 2018MNRAS.478.5533F,2015MNRAS.452.3838C}. While the inclusion of fainter satellites would constitute a more detailed comparison to cosmological expectations, current state-of-the-art simulations of sufficiently large volumes to contain a large number of MW-like hosts do not simultaneously resolve a sufficient number of lower-mass satellite galaxies to allow such a comparison. At the same time, the observed sample of fainter satellite galaxies suffers from various, difficult to quantify biases due to the footprints and incompletness limits of the different surveys in which they were discovered \citep[e.g.][]{2016MNRAS.456..448P}.

\subsection{The Proper Motion Data}
\label{sec:PMdata}

\begin{table*}
	\centering
	\caption{The adopted data for the classical MW satellite galaxies. The heliocentric galaxy positions are given in Galactic longitude $l$, latitude $b$\ and heliocentric distance $r_{\sun}$, taken from \citet{2012AJ....144....4M}. The heliocentric velocities are given as the line-of-sight velocity $v_{\mathrm{los}}$\ relative to the Sun and the two PM components $\mu_{\alpha} \cos \delta$\ and $\mu_{\delta}$. For the Gaia PMs, also the correlation coefficients $\mathrm{C}_{\mu_{\alpha}, \mu_{\delta}}$\ and the systematic errors $\epsilon_\mathrm{sys}$\ are provided. Type indicates whether the PM is determined from ground-based observations (Ground), from Hubble Space Telescope observations (HST), from Gaia DR2 (Gaia), or via the stellar redshift gradient method (SRG). Proper motions that were used for the Combined sample (see Sect. \ref{sec:PMdata}) are marked with a $^\dagger$. The last column gives the reference for the respective PM measurement.}
	\label{tab:satellitedata}
	\scriptsize
	\begin{tabular}{lcccccccccl}
  \hline
  Name & $l$ & $b$ & $r_{\sun}$ & $v_{\mathrm{los}}$ & $\mu_{\alpha} \cos \delta$ & $\mu_{\delta}$ & $\mathrm{C}_{\mu_{\alpha}, \mu_{\delta}}$ & $\epsilon_\mathrm{sys}$ & Type & Ref. \\
   & $[^\circ]$ & $[^\circ]$ & [kpc] & $[\mathrm{km\,s}^{-1}]$ & $[\mathrm{mas\,yr}^{-1}]$ & $[\mathrm{mas\,yr}^{-1}]$ &  & $[\mathrm{mas\,yr}^{-1}]$ &  &  \\
   \hline
Sgr &  5.6 & -14.2 & 26.3 & $140.0 \pm 2.0$ & $-2.650 \pm 0.080$ &  $-0.880 \pm 0.080$ &       &    & Ground$^\dagger$ & \citet{1997AJ....113..634I} \\
 & & & & &  $-2.830 \pm 0.200$ &  $-1.330 \pm 0.200$ &     &    & Ground & \citet{2005ApJ...618L..25D} \\
 & & & & &  $-2.750 \pm 0.200$ &  $-1.650 \pm 0.220$ &       &    & HST & \citet{2010AJ....139..839P} \\
 & & & & &  $-2.692 \pm 0.001$ &  $-1.359 \pm 0.001$ &  +0.090 & 0.035 & Gaia$^\dagger$ & \citet{2018AA...616A..12G} \\
 & & & & &  $-2.736 \pm 0.009$ &  $-1.357 \pm 0.008$ &  +0.114 & 0.035 & Gaia & \citet{2018AA...619A.103F} \\
LMC &  280.5 & -32.9 & 50.6 & $262.2 \pm 3.4$ & $+2.030 \pm 0.080$ &  $+0.440 \pm 0.050$ &       &    & HST & \citet{2006ApJ...638..772K} \\
 & & & & &  $+1.956 \pm 0.036$ &  $+0.435 \pm 0.036$ &       &    & HST & \citet{2008AJ....135.1024P} \\
 & & & & &  $+1.910 \pm 0.020$ &  $+0.229 \pm 0.047$ &       &    & HST$^\dagger$ & \citet{2013ApJ...764..161K} \\
 & & & & &  $+1.850 \pm 0.010$ &  $+0.234 \pm 0.010$ &     & 0.028 & Gaia$^\dagger$ & \citet{2018AA...616A..12G} \\
SMC &  302.8 & -44.3 & 64.0 & $145.6 \pm 0.6$ & $+1.160 \pm 0.180$ &  $-1.170 \pm 0.180$ &       &    & HST & \citet{2006ApJ...652.1213K} \\
 & & & & &  $+0.754 \pm 0.061$ &  $-1.252 \pm 0.058$ &       &    & HST & \citet{2008AJ....135.1024P} \\
 & & & & &  $+0.772 \pm 0.063$ &  $-1.117 \pm 0.061$ &       &    & HST$^\dagger$ & \citet{2013ApJ...764..161K} \\
 & & & & &  $+0.797 \pm 0.010$ &  $-1.220 \pm 0.010$ &     & 0.028 & Gaia$^\dagger$ & \citet{2018AA...616A..12G} \\
Dra &  86.4 & 34.7 & 75.9 & $-291.0 \pm 0.1$ & $+0.600 \pm 0.400$ &  $+1.100 \pm 0.500$ &       &    & Ground & \citet{1994IAUS..161..535S} \\
 & & & & &  $+0.177 \pm 0.063$ &  $-0.221 \pm 0.063$ &       &    & HST & \citet{2015AJ....149...42P} \\
 & & & & &  $-0.284 \pm 0.047$ &  $-0.289 \pm 0.041$ &       &    & Ground$^\dagger$ & \citet{2016MNRAS.461..271C} \\
 & & & & &  $+0.056 \pm 0.010$ &  $-0.176 \pm 0.010$ &       &    & HST$^\dagger$ & \citet{2017ApJ...849...93S} \\
 & & & & &  $-0.019 \pm 0.009$ &  $-0.145 \pm 0.010$ &  -0.080 & 0.035 & Gaia$^\dagger$ & \citet{2018AA...616A..12G} \\
 & & & & &  $-0.013 \pm 0.013$ &  $-0.158 \pm 0.015$ &  +0.131 & 0.035 & Gaia & \citet{2018AA...619A.103F} \\
UMi &  105.0 & 44.8 & 75.9 & $-246.9 \pm 0.1$ & $+0.500 \pm 0.800$ &  $+1.200 \pm 0.500$ &       &    & Ground & \citet{1994IAUS..161..535S} \\
 & & & & &  $+0.056 \pm 0.078$ &  $+0.074 \pm 0.099$ &       &    & Ground$^\dagger$ & \citet{1997ASPC..127..103S} \\
 & & & & &  $-0.500 \pm 0.170$ &  $+0.220 \pm 0.160$ &       &    & HST & \citet{2005AJ....130...95P} \\
 & & & & &  $-0.182 \pm 0.010$ &  $+0.074 \pm 0.008$ &  -0.340 & 0.035 & Gaia$^\dagger$ & \citet{2018AA...616A..12G} \\
 & & & & &  $-0.184 \pm 0.026$ &  $+0.082 \pm 0.023$ &  -0.387 & 0.035 & Gaia & \citet{2018AA...619A.103F} \\
Scl &  287.5 & -83.2 & 85.9 & $111.4 \pm 0.1$ & $+0.720 \pm 0.220$ &  $-0.060 \pm 0.250$ &       &    & Ground & \citet{1995AJ....110.2747S} \\
 & & & & &  $+0.090 \pm 0.130$ &  $+0.020 \pm 0.130$ &       &    & HST & \citet{2006AJ....131.1445P} \\
 & & & & &  $-0.400 \pm 0.290$ &  $-0.690 \pm 0.470$ &       &    & SRG & \citet{2008ApJ...688L..75W} \\
 & & & & &  $+0.030 \pm 0.021$ &  $-0.136 \pm 0.021$ &       &    & HST$^\dagger$ & \citet{2017ApJ...849...93S} \\
 & & & & &  $+0.082 \pm 0.005$ &  $-0.131 \pm 0.004$ &  +0.230 & 0.035 & Gaia$^\dagger$ & \citet{2018AA...616A..12G} \\
 & & & & &  $+0.085 \pm 0.006$ &  $-0.133 \pm 0.006$ &  +0.157 & 0.035 & Gaia & \citet{2018AA...619A.103F} \\
Sxt &  243.5 & 42.3 & 85.9 & $224.2 \pm 0.1$ & $-0.260 \pm 0.410$ &  $+0.100 \pm 0.440$ &       &    & SRG & \citet{2008ApJ...688L..75W} \\
 & & & & &  $-0.409 \pm 0.050$ &  $-0.047 \pm 0.058$ &       &    & Ground$^\dagger$ & \citet{2018MNRAS.473.4064C} \\
 & & & & &  $-0.496 \pm 0.025$ &  $+0.077 \pm 0.020$ &  -0.450 & 0.035 & Gaia$^\dagger$ & \citet{2018AA...616A..12G} \\
 & & & & &  $-0.438 \pm 0.028$ &  $+0.055 \pm 0.028$ &  -0.238 & 0.035 & Gaia & \citet{2018AA...619A.103F} \\
Car &  260.1 & -22.2 & 105.2 & $222.9 \pm 0.1$ & $+0.220 \pm 0.090$ &  $+0.150 \pm 0.090$ &       &    & HST$^\dagger$ & \citet{2003AJ....126.2346P} \\
 & & & & &  $+0.250 \pm 0.360$ &  $+0.160 \pm 0.430$ &       &    & SRG & \citet{2008ApJ...688L..75W} \\
 & & & & &  $+0.495 \pm 0.015$ &  $+0.143 \pm 0.014$ &  -0.080 & 0.035 & Gaia$^\dagger$ & \citet{2018AA...616A..12G} \\
 & & & & &  $+0.485 \pm 0.018$ &  $+0.132 \pm 0.016$ &  +0.083 & 0.035 & Gaia & \citet{2018AA...619A.103F} \\
Fnx &  237.1 & -65.7 & 147.2 & $55.3 \pm 0.1$ & $+0.590 \pm 0.160$ &  $-0.150 \pm 0.160$ &       &    & HST & \citet{2004AJ....128..687D} \\
 & & & & &  $+0.476 \pm 0.046$ &  $-0.360 \pm 0.041$ &       &    & HST$^\dagger$ & \citet{2007AJ....133..818P} \\
 & & & & &  $+0.480 \pm 0.150$ &  $-0.250 \pm 0.140$ &       &    & SRG & \citet{2008ApJ...688L..75W} \\
 & & & & &  $+0.620 \pm 0.160$ &  $-0.530 \pm 0.150$ &       &    & Ground & \citet{2011AJ....142...93M} \\
 & & & & &  $+0.376 \pm 0.003$ &  $-0.413 \pm 0.003$ &  -0.090 & 0.035 & Gaia$^\dagger$ & \citet{2018AA...616A..12G} \\
 & & & & &  $+0.375 \pm 0.004$ &  $-0.401 \pm 0.005$ &  -0.460 & 0.035 & Gaia & \citet{2018AA...619A.103F} \\
Leo\,II &  220.2 & 67.2 & 233.3 & $78.0 \pm 0.1$ & $+0.104 \pm 0.113$ &  $-0.033 \pm 0.151$ &       &    & HST & \citet{2011ApJ...741..100L} \\
 & & & & &  $-0.069 \pm 0.037$ &  $-0.087 \pm 0.039$ &       &    & HST$^\dagger$ & \citet{2016AJ....152..166P} \\
 & & & & &  $-0.064 \pm 0.057$ &  $-0.210 \pm 0.054$ &  +0.050 & 0.035 & Gaia$^\dagger$ & \citet{2018AA...616A..12G} \\
 & & & & &  $+0.020 \pm 0.090$ &  $-0.201 \pm 0.093$ &  -0.401 & 0.035 & Gaia & \citet{2018AA...619A.103F} \\
Leo\,I &  226.0 & 49.1 & 253.5 & $282.5 \pm 0.1$ & $-0.114 \pm 0.029$ &  $-0.126 \pm 0.029$ &       &    & HST$^\dagger$ & \citet{2013ApJ...768..139S} \\
 & & & & &  $-0.097 \pm 0.056$ &  $-0.091 \pm 0.047$ &  -0.510 & 0.035 & Gaia$^\dagger$ & \citet{2018AA...616A..12G} \\
 & & & & &  $-0.086 \pm 0.059$ &  $-0.128 \pm 0.062$ &  -0.358 & 0.035 & Gaia & \citet{2018AA...619A.103F} \\
  \hline
	\end{tabular}
\end{table*}

The positions, distances, and line-of-sight velocities to the 11 classical MW satellite galaxies are compiled in Table \ref{tab:satellitedata}. These were adopted from \citet{2013MNRAS.435.2116P}. The table also lists all recent proper motion measurements for these satellite galaxies and the corresponding references. To investigate the effect Gaia DR2 data has on the deduced distribution of satellite galaxy orbital poles, we investigate three samples:

\begin{itemize}
\item {\it Pre-Gaia 2018}: Following the approach taken in \citet{2013MNRAS.435.2116P}, this sample uses an uncertainty-weighted average of the best-available proper motions for each of the classical satellite galaxies available before the publication of Gaia DR2.
\item {\it Gaia DR2 Only}: This sample exclusively considers proper motions from Gaia DR2, making it completely independent from the previous sample. We opt for using the Gaia DR2 proper motions by \citet{2018AA...616A..12G} over those of \citet{2018AA...619A.103F} because the former report slightly smaller uncertainties.
\item {\it Combined}: For this sample, the Gaia DR2 proper motions reported by \citet{2018AA...616A..12G} have been combined with the best-available proper motion measurement from a different technique, i.e. the measurements with the smallest reported uncertainties (marked with $^\dagger$\ in Table \ref{tab:satellitedata}). Each satellite is assigned their uncertainty-weighted average. Effectively, this means that most satellites are assigned the Gaia DR2 proper motions, except Leo\,I and II for which much higher precision HST data is available from \citet{2016AJ....152..166P} and \cite{2013ApJ...768..139S}. This is our fiducial sample and represents our best current knowledge of the proper motions of the 11 classical satellite galaxies of the Milky Way.
\end{itemize}

In addition, we compare to the results \citet{2013MNRAS.435.2116P} obtained with the then-available, in part much less accurate, proper motions.

\subsection{Determining the Orbital Poles}
\label{sec:OrbPoles}

The Cartesian positions and velocities, orbital poles and specific angular momenta around the MW center for each of the satellites are calculated as described in \citet{2013MNRAS.435.2116P}.  In short, we convert the heliocentric satellite positions, velocities, and proper motions to a Cartesian Galactic coordinate system centered on the Galactic Center, calculate the angular momentum vector as the cross product of the position and velocity vectors of the satellites, and determine its direction (i.e. the orbital pole in Galactic coordinates $l_\mathrm{pole}$\ and $b_\mathrm{pole}$) and length (i.e. the specific angular momentum $h$). To estimate the uncertainties, we generate 10\,000 realizations by drawing from the uncertainties of the satellite positions, velocities, and proper motions. For the Gaia DR2 sample we take into account that some correlation between the two proper motion components exists.
Our procedure assumes a local standard of rest (LSR) with a circular velocity of $v_\mathrm{LSR} = 239\,\mathrm{km\,s}^{-1}$ \citep{2011MNRAS.414.2446M}, a peculiar velocity of the Sun relative to the LSR of $(U, V, W) = (11.10\,\mathrm{km\,s}^{-1}, 12.24\,\mathrm{km\,s}^{-1}, 7.25\,\mathrm{km\,s}^{-1})$ \citep{2010MNRAS.403.1829S}, and a distance of $d_{\sun} = 8.3\,\mathrm{kpc}$\ from the Sun to the Galactic Center \citep{2011MNRAS.414.2446M} to be fully comparable to \citet{2013MNRAS.435.2116P}.

\subsection{Results}

\begin{table*}
	\centering
	\caption{Orbital poles of the 11 classical MW satellites, using the {\it combined} PMs. 
 The three position ($x, y, z$) and velocity ($v_{\mathrm{x}}, v_{\mathrm{y}}, v_{\mathrm{z}}$) components are given in Galactic Cartesian coordinates. The orbital pole direction is expressed in Galactic longitude, $l_{\mathrm{pole}}$, and Galactic latitude, $b_{\mathrm{pole}}$. The orbital pole uncertainty $\Delta_{\mathrm{pole}}$\ is oriented along the direction of the great circle perpendicular to the position of the satellite galaxy. 
The absolute specific angular momentum is $h = \left| \mathbf{r} \times \mathbf{v} \right|$. It is negative for those satellites which counter-orbit relative to the average orbital pole of the MW satellites (as determined from the 7 most-concentrated poles). Also given are the angular distance of the orbital poles from normal to the plane fitted to all 11 classical satellites, $\theta_{\mathrm{VPOS}}^{\mathrm{class}}$, and the angle between the orbital poles and the average direction of the six most-concentrated orbital poles,  $\theta_{\mathrm{J6}}$.
	}
	\label{tab:individualpoles}
	\begin{tabular}{lcccccccccccc}
  \hline
  Name & $x$ & $y$ & $z$ & $v_{\mathrm{x}} $ & $v_{\mathrm{y}} $  & $v_{\mathrm{z}} $ & $l_{\mathrm{pole}} $ & $b_{\mathrm{pole}} $ & $\Delta_{\mathrm{pole}} $ & $h / 10^3$ & $\theta_{\mathrm{VPOS}}^{\mathrm{class}}$ & $\theta_{\mathrm{J6}} $ \\
 & kpc & kpc & kpc & $[\mathrm{km\,s}^{-1}]$ & $[\mathrm{km\,s}^{-1}]$  & $[\mathrm{km\,s}^{-1}]$ & $[^{\circ}]$ & $[^{\circ}]$ & $[^{\circ}]$ & $[\mathrm{kpc\,km\,s^{-1}}]$ & $[^{\circ}]$ & $[^{\circ}]$ \\
  \hline
Sagittarius &  17.1 & 2.5 & -6.4 & $233 \pm  2$ & $ -8 \pm  4$ & $209 \pm  4$ & 275.2 & -8.0 & 0.8 & $-5.1 \pm 0.1$ & $ 118.2_{-0.4}^{+0.4} $ &  $93.3$\\
LMC &  -0.6 & -41.8 & -27.5 & $-42 \pm  6$ & $-223 \pm  4$ & $231 \pm  4$ & 175.3 & -5.9 & 1.1 & $15.9 \pm 0.2$ & $ 19.2_{-0.4}^{+0.4} $ &  $8.5$\\
SMC &  16.5 & -38.5 & -44.7 & $  6 \pm  8$ & $-180 \pm  7$ & $167 \pm  6$ & 191.8 & -10.5 & 1.9 & $15.0 \pm 0.5$ & $ 36.1_{-1.1}^{+1.2} $ &  $11.9$\\
Draco &  -4.3 & 62.2 & 43.2 & $ 62 \pm  3$ & $ 14 \pm  2$ & $-166 \pm  3$ & 169.9 & -19.3 & 1.2 & $11.8 \pm 0.3$ & $ 21.6_{-0.3}^{+0.3} $ &  $11.5$\\
Ursa Minor &  -22.2 & 52.0 & 53.5 & $  7 \pm 12$ & $ 56 \pm  9$ & $-154 \pm  9$ & 195.5 & -8.0 & 4.3 & $11.6 \pm 0.9$ & $ 39.3_{-2.5}^{+2.7} $ &  $16.1$\\
Sculptor &  -5.2 & -9.8 & -85.3 & $ 31 \pm  7$ & $184 \pm  7$ & $-97 \pm  1$ & 349.3 & -2.2 & 2.0 & $-16.9 \pm 0.6$ & $ 166.4_{-1.9}^{+1.9} $ &  $161.4$\\
Sextans &  -36.7 & -56.9 & 57.8 & $-221 \pm 13$ & $ 81 \pm 10$ & $ 59 \pm 10$ & 232.7 & -49.4 & 3.4 & $20.5 \pm 1.2$ & $ 79.5_{-3.4}^{+3.4} $ &  $56.3$\\
Carina &  -25.0 & -95.9 & -39.8 & $-46 \pm 16$ & $-39 \pm  7$ & $143 \pm 16$ & 160.5 & -11.9 & 6.5 & $16.7 \pm 1.8$ & $ 10.5_{-3.0}^{+5.3} $ &  $19.0$\\
Fornax &  -41.3 & -51.0 & -134.1 & $ 17 \pm 18$ & $-140 \pm 18$ & $ 94 \pm  8$ & 176.5 & 15.8 & 6.7 & $24.7 \pm 2.8$ & $ 26.8_{-6.2}^{+6.4} $ &  $29.1$\\
Leo II &  -77.3 & -58.3 & 215.2 & $-24 \pm 34$ & $ 86 \pm 36$ & $ 36 \pm 14$ & 186.3 & -21.2 & 23.2 & $24.0 \pm 8.6$ & $ 34.8_{-20.7}^{+22.5} $ &  $10.1$\\
Leo I &  -123.6 & -119.3 & 191.7 & $-167 \pm 29$ & $-28 \pm 28$ & $100 \pm 21$ & 251.1 & -38.6 & 17.9 & $28.1 \pm 8.0$ & $ 92.3_{-17.4}^{+17.4} $ &  $67.2$\\
  \hline
	\end{tabular}
\end{table*}

Table \ref{tab:individualpoles} lists the results for the Combined sample. It contains the Cartesian, Galactocentric positions of the satellites ($x, y, z$), their 3D Cartesian velocities ($v_\mathrm{x} ,v_\mathrm{y}, v_\mathrm{z} $), the directions of their most-likely orbital poles in Galactic longitude $l_\mathrm{pole}$ and latitude $b_\mathrm{pole}$, the uncertainties in this direction $\Delta_\mathrm{pole}$ (along a great-circle segment perpendicular to the line connecting the satellite and the Galactic center), and the specific angular momentum $h$\ of the satellite. Also given is the angle $\theta_\mathrm{VPOS}^\mathrm{class}$\ between the normal vector to the VPOSclass plane, i.e. the plane minimizing the distance to the 11 classical satellite galaxies. The VPOSclass normal vector points to $(l, b) = (157^\circ.3, -12^\circ.7)$\ in Galactic coordinates \citep{2007MNRAS.374.1125M}. Finally, the table lists the angle to the average direction of the six most-concentrated orbital poles, $\theta_\mathrm{J6}$.

For each of the three samples (Sect. \ref{sec:PMdata}), we measure the average direction and concentration $\Delta_\mathrm{std}(k)$\ of the $k = [3, ..., 11]$\ most closely aligned orbital poles with the following procedure: For a given $k$, we determine every possible combination of $k$\ satellites. For each of these combinations, we calculate the average direction $\langle\boldsymbol{n}\rangle$\ of their orbital poles on the unit sphere and measure the spherical standard distance of these orbital poles,

\begin{equation}
\Delta_{\mathrm{sph}}(k) = \sqrt{\frac{\sum_{i=1}^{k}\left[\arccos \left(\langle\boldsymbol{n}\rangle \cdot \boldsymbol{n}_{i}\right)\right]^{2}}{k}},
\label{eq:Delta_std}
\end{equation}
where $\boldsymbol{n}_{i}$\ are the individual orbital pole directions \citep{2007MNRAS.374.1125M, 2013MNRAS.435.1928P, 2013MNRAS.435.2116P}. The sample resulting in the smallest $\Delta_\mathrm{std}$\ for a given $k$\ is chosen, and the spherical standard distance as well as the corresponding average direction of these orbital poles is recorded.

\begin{figure*}
	\includegraphics[width=140mm]{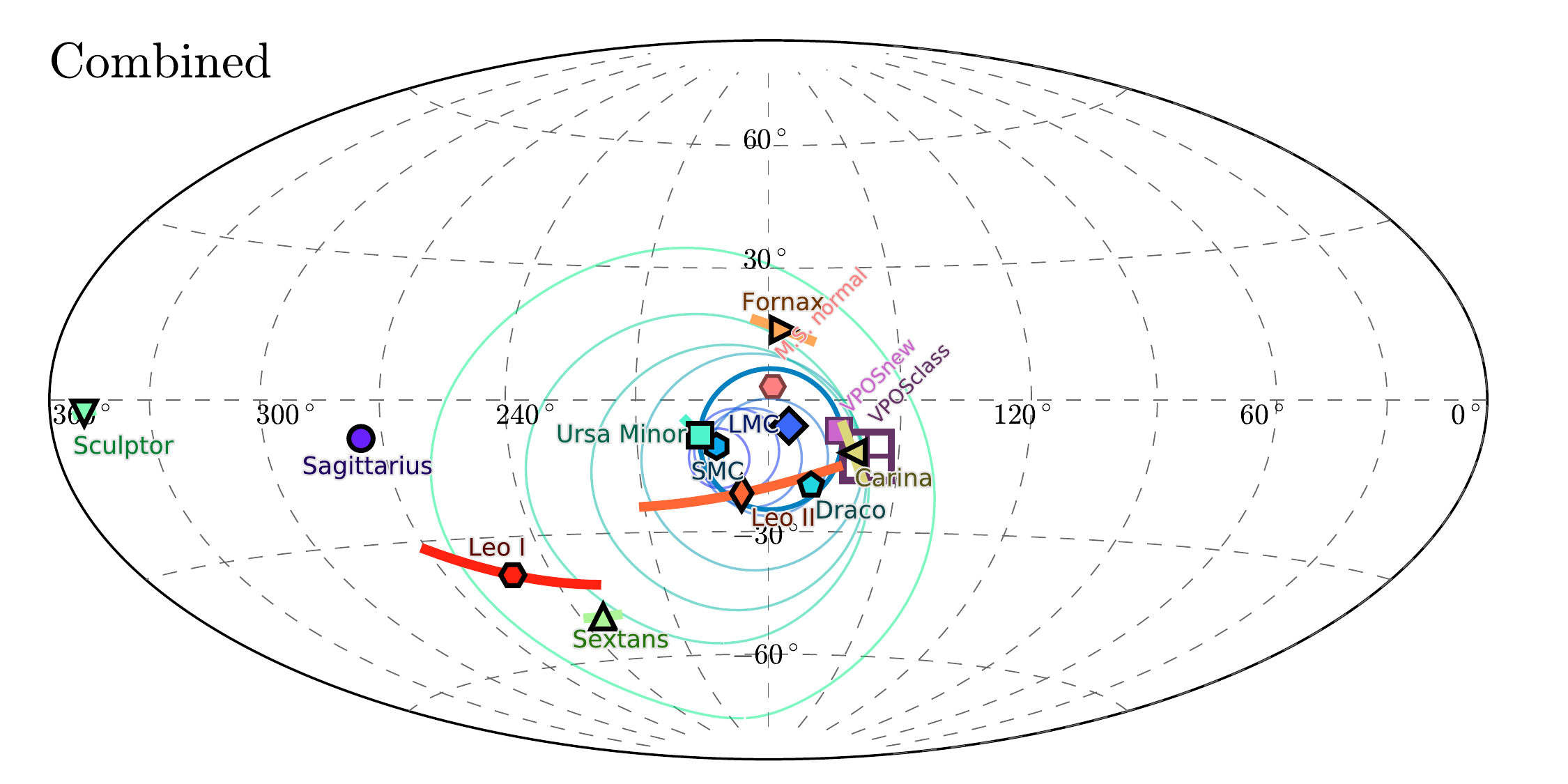}
	\includegraphics[width=140mm]{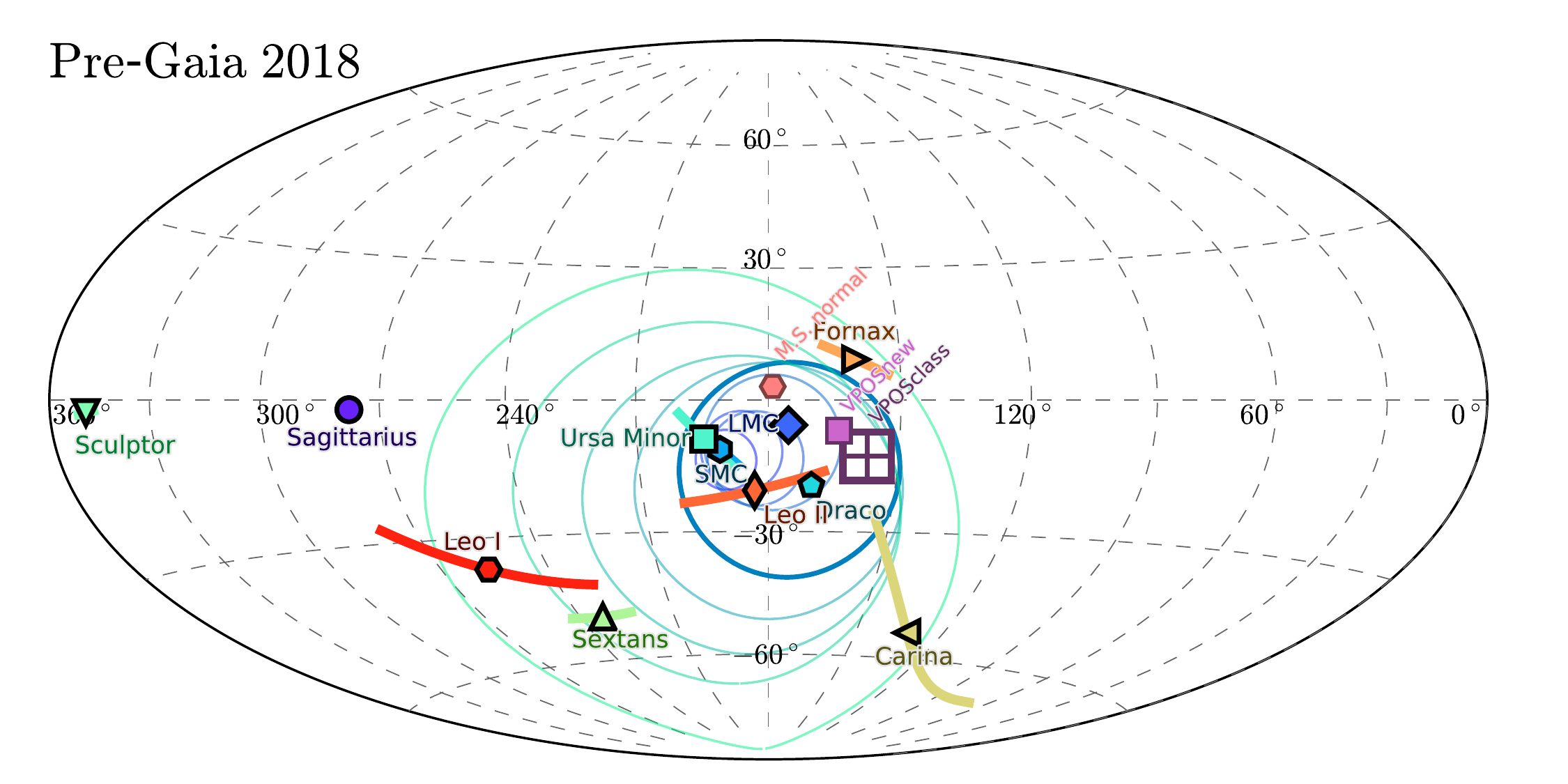}
	\includegraphics[width=140mm]{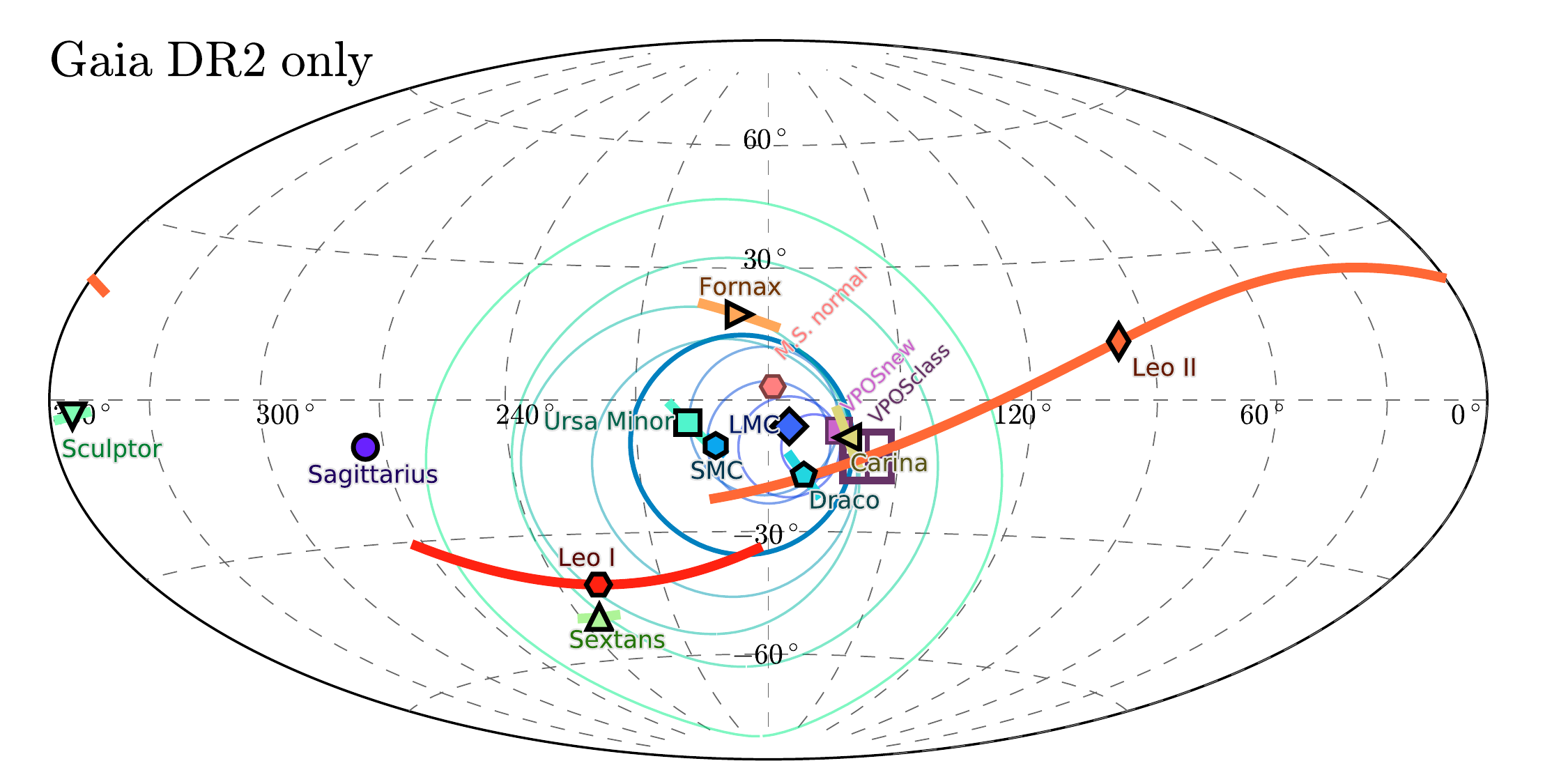}
    \caption{
      Orbital pole distribution in Galactic coordinates, for three PM samples: the combined sample of Gaia DR2 and the best-available HST measurements (top), the combination of the best available PMs prior to Gaia DR2 (middle), and using only the Gaia DR2 PMs (bottom) as reported by \citet{2018AA...616A..12G}. Uncertainties in the orbital pole directions are indicated by the great-circle segments. The circles indicate the location, and their radii the spherical standard distance $\Delta_{\mathrm{std}}$, of the $k = [3, ..., 11]$\ most concentrated orbital poles (increasing radii correspond to increasing $k$). The circle for $k = 7$ is emphasized, colors follow the same scheme as in the top panel of Fig. \ref{fig:history}. Also indicated is the direction of the normal vectors to the plane fitted to the positions of the 11 classical satellites (VPOSclass, from \citealt{2007MNRAS.374.1125M}), and to the positions of the full sample of satellites in the VPOS (VPOSnew, from \citet{2015MNRAS.453.1047P}). Finally, the normal to the Magellanic Stream Plane (M.S. normal) is given, as defined in \citet{2012MNRAS.423.1109P}.
    }
    \label{fig:asps}
\end{figure*}

\begin{table*}
	\centering
	\caption{Direction $\langle\boldsymbol{n}\rangle$ of the average orbital pole of the $k$\ most concentrated orbital poles in Galactic longitude $l_\mathrm{av}$\ and latitude $b_\mathrm{av}$, angle $\theta_{\mathrm{VPOS}}$ between this direction and the VPOSclass normal vector of the plane minimizing the distances to the 11 classical satellites, and the spherical standard distances $\Delta_{\mathrm{sph}}(k)$\ of the $k$ most-concentrated orbital poles for four samples of PM measurements: as reported by \citet{2013MNRAS.435.2116P}, for the best PM measurements available prior to Gaia DR2, using only Gaia DR2 PMs, and for the Combined sample. For each the frequencies of finding as or more concentrated orbital poles for random velocities ($f_\mathrm{random}$) and for satellite systems in the Illustris TNG100 cosmological simulation ($f_\mathrm{TNG100}$\ for the hydrodynamical run, $f_\mathrm{TNG100-Dark}$\ for its dark-matter-only equivalent) are also given. For the Combined sample, the fraction of systems in TNG100-1 that fulfill both the orbital pole concentration criterion used for $f_\mathrm{TNG100}$\ as well as one of two flattening criteria ($c/a \leq 0.182$\ for $f_{\Delta_\mathrm{std} + c/a}^\mathrm{TNG100}$\ or $r_\mathrm{per} \leq 19.6\,\mathrm{kpc}$\ for $f_{\Delta_\mathrm{std} + r_\mathrm{per}}^\mathrm{TNG100}$) are also given.}
	\label{tab:deltasph}
	\begin{tabular}{lccccccccc}
  \hline
  $k =$ & 3 & 4 & 5 & 6 & 7 & 8 & 9 & 10 & 11 \\
   \hline
   Pawlowski \& Kroupa 2013 & & & & & & & & & \\
~~$\langle\boldsymbol{n}\rangle \begin{pmatrix} l_\mathrm{av} \\ b_\mathrm{av} \end{pmatrix}[^\circ]$ & 
$\begin{pmatrix} 187.0 \\ -8.7 \end{pmatrix}$ & 
$\begin{pmatrix} 187.9 \\ -4.2 \end{pmatrix}$ & 
$\begin{pmatrix} 182.4 \\ -1.5 \end{pmatrix}$ & 
$\begin{pmatrix} 177.3 \\ -3.2 \end{pmatrix}$ & 
$\begin{pmatrix} 180.1 \\ -9.0 \end{pmatrix}$ & 
$\begin{pmatrix} 176.4 \\ -15.0 \end{pmatrix}$ & 
$\begin{pmatrix} 182.9 \\ -19.2 \end{pmatrix}$ & 
$\begin{pmatrix} 191.1 \\ -19.5 \end{pmatrix}$ & 
$\begin{pmatrix} 191.7 \\ -23.1 \end{pmatrix}$  \\
~~$\Delta_{\mathrm{sph}}(k)\ [^\circ]$ &  8.5 & 10.8 & 15.6 & 18.5 & 23.6 & 29.3 & 35.9 & 44.1 & 62.3 \\
~~$f_\mathrm{random}\ [\%]$ & 9.52 & 1.48 & 0.92 & 0.16 & 0.12 & 0.12 & 0.32 & 0.52 & 2.84 \\
~~$f_\mathrm{TNG100-Dark}\ [\%]$  & 15.34 & 5.38 & 4.17 & 2.03 & 1.52 & 1.56 & 1.52 & 1.21 & 7.88 \\
~~$f_\mathrm{TNG100}\ [\%]$ & 17.50 & 5.30 & 4.67 & 2.20 & 2.04 & 1.81 & 1.65 & 1.77 & 8.20 \\
  \hline
Pre-Gaia 2018 & & & & & & & & \\
~~$\langle\boldsymbol{n}\rangle \begin{pmatrix} l_\mathrm{av} \\ b_\mathrm{av} \end{pmatrix}[^\circ]$ & 
$\begin{pmatrix} 189.7 \\ -13.6 \end{pmatrix}$ & 
$\begin{pmatrix} 186.1 \\ -11.7 \end{pmatrix}$ & 
$\begin{pmatrix} 182.9 \\ -13.3 \end{pmatrix}$ & 
$\begin{pmatrix} 179.1 \\ -9.7 \end{pmatrix}$ & 
$\begin{pmatrix} 175.0 \\ -16.2 \end{pmatrix}$ & 
$\begin{pmatrix} 179.9 \\ -21.3 \end{pmatrix}$ & 
$\begin{pmatrix} 186.6 \\ -25.0 \end{pmatrix}$ & 
$\begin{pmatrix} 195.0 \\ -25.1 \end{pmatrix}$ & 
$\begin{pmatrix} 199.3 \\ -28.6 \end{pmatrix}$  \\
~~$\Delta_{\mathrm{sph}}(k)\ [^\circ]$ &  6.8 & 9.1 & 10.7 & 15.4 & 24.7 & 29.7 & 35.0 & 42.7 & 58.1 \\
~~$f_\mathrm{random}\ [\%]$ & 4.0 & 0.72 & 0.08 & 0.08 & 0.20 & 0.12 & 0.16 & 0.36 & 1.12 \\
~~$f_\mathrm{TNG100-Dark}\ [\%]$  & 7.25 & 2.57 & 0.59 & 0.59 & 1.76 & 1.64 & 1.29 & 0.86 & 3.90 \\
~~$f_\mathrm{TNG100}\ [\%]$ & 9.81 & 2.75 & 0.75 & 0.663 & 2.79 & 1.92 & 1.41 & 1.30 & 4.40 \\
  \hline
Gaia DR2 only & & & & & & & & \\
~~$\langle\boldsymbol{n}\rangle \begin{pmatrix} l_\mathrm{av} \\ b_\mathrm{av} \end{pmatrix}[^\circ]$ & 
$\begin{pmatrix} 169.6 \\ -10.6 \end{pmatrix}$ & 
$\begin{pmatrix} 175.2 \\ -10.7 \end{pmatrix}$ & 
$\begin{pmatrix} 179.9 \\ -9.6 \end{pmatrix}$ & 
$\begin{pmatrix} 181.0 \\ -4.8 \end{pmatrix}$ & 
$\begin{pmatrix} 186.1 \\ -10.4 \end{pmatrix}$ & 
$\begin{pmatrix} 190.2 \\ -15.9 \end{pmatrix}$ & 
$\begin{pmatrix} 198.4 \\ -16.8 \end{pmatrix}$ & 
$\begin{pmatrix} 190.2 \\ -15.4 \end{pmatrix}$ & 
$\begin{pmatrix} 193.2 \\ -18.2 \end{pmatrix}$  \\
~~$\Delta_{\mathrm{sph}}(k)\ [^\circ]$ &  7.3 & 11.4 & 13.9 & 16.8 & 25.1 & 29.7 & 37.9 & 47.8 & 64.6 \\
~~$f_\mathrm{random}\ [\%]$ & 5.32 & 1.96 & 0.64 & 0.08 & 0.32 & 0.12 & 0.52 & 0.96 & 4.6 \\
~~$f_\mathrm{TNG100-Dark}\ [\%]$  & 9.40 & 6.40 & 2.46 & 1.17 & 2.11 & 1.64 & 2.34 & 3.00 & 11.12 \\
~~$f_\mathrm{TNG100}\ [\%]$ & 11.50 & 6.51 & 2.71 & 1.30 & 3.06 & 1.92 & 2.32 & 3.41 & 12.44 \\
  \hline
Combined & & & & & & & & & \\
~~$\langle\boldsymbol{n}\rangle \begin{pmatrix} l_\mathrm{av} \\ b_\mathrm{av} \end{pmatrix}[^\circ]$ & 
$\begin{pmatrix} 191.3 \\ -13.2 \end{pmatrix}$ & 
$\begin{pmatrix} 187.3 \\ -11.5 \end{pmatrix}$ & 
$\begin{pmatrix} 183.9 \\ -13.1 \end{pmatrix}$ & 
$\begin{pmatrix} 180.0 \\ -13.1 \end{pmatrix}$ & 
$\begin{pmatrix} 179.5 \\ -9.0 \end{pmatrix}$ & 
$\begin{pmatrix} 183.7 \\ -14.4 \end{pmatrix}$ & 
$\begin{pmatrix} 189.3 \\ -18.2 \end{pmatrix}$ & 
$\begin{pmatrix} 196.9 \\ -18.9 \end{pmatrix}$ & 
$\begin{pmatrix} 200.8 \\ -21.4 \end{pmatrix}$  \\
~~$\theta_{\mathrm{VPOSclass}}(k)\ [^\circ]$ & 33.1 & 29.3 & 25.9 & 22.1 & 22.1 & 25.7 & 31.3 & 38.5 & 42.3\\
~~$\theta_{\mathrm{VPOSnew}}(k)\ [^\circ]$ & 27.6 & 23.4 & 25.9 & 20.5 & 16.9 & 15.5 & 20.7 & 27.1 & 38.4\\
~~$\Delta_{\mathrm{sph}}(k)\ [^\circ]$ &  6.8 & 9.5 & 11.3 & 13.3 & 16.0 & 24.8 & 30.6 & 38.2 & 56.0 \\
~~$f_\mathrm{random}\ [\%]$  & 4.0 & 0.80 & 0.08 & 0.04 & 0.0 & 0.0 & 0.04 & 0.04 & 0.64 \\
~~$f_\mathrm{bestaligned}\ [\%]$  & 1.24 & 0.28 & 1.24 & 2.40 & 0.44 & 0.0 & 0.0 & 0.0 & 0.60 \\ 
~~$f_{\Delta_\mathrm{std}}^\mathrm{TNG100-Dark}\ [\%]$  & 7.25 & 3.12 & 0.82 & 0.31 & 0.04 & 0.51 & 0.39 & 0.43 & 2.15 \\
~~$f_{\Delta_\mathrm{std}}^\mathrm{TNG100}\ [\%]$  & 9.81 & 3.22 & 1.02 & 0.39 & 0.04 & 0.63 & 0.55 & 0.59 & 3.10 \\
~~$f_{\Delta_\mathrm{std} + c/a}^\mathrm{TNG100}\ [\%]$ & 0.20 & 0.08 & 0.04 & 0.04 & 0.0 & 0.08 & 0.0 & 0.0 & 0.0\\
~~$f_{\Delta_\mathrm{std} + r_\mathrm{per}}^\mathrm{TNG100}\ [\%]$ & 0.08 & 0.0 & 0.0 & 0.0 & 0.0 & 0.0 & 0.0 & 0.0 & 0.0\\
  \hline
	\end{tabular}
\end{table*}

The distribution of orbital poles and their uncertainties (green great-circle segments) for our three samples is shown in Figure \ref{fig:asps}. These plots also indicate three directions derived from spatially constrained structures:
\begin{itemize}
\item VPOSclass: the minor axis direction for the distribution of the 11 classical satellites, pointing to $(l, b) = (157^\circ.3, -12^\circ.7)$\ \citep{2007MNRAS.374.1125M},
\item VPOSnew:  the minor axis direction of the overall Milky Way satellite system, pointing to $(l, b) = (164^\circ.0, -6^\circ.9)$\ \citep{2015MNRAS.453.1047P}, and
\item M.S. normal: the Magellanic Stream normal vector, pointing to $(l, b) = (179^\circ, 3^\circ)$ \citep{2012MNRAS.423.1109P}. 
\end{itemize}

Table  \ref{tab:deltasph} compiles the average directions $\langle\boldsymbol{n}\rangle$\ of the $k = 3$\ to 11 most concentrated orbital poles (in Galactic coordinates $l_\mathrm{av}$\ and $b_\mathrm{av}$), the angle $\theta_{\mathrm{VPOS}}(k)$\ between this direction and the VPOSclass normal vector, and the spherical standard distance $\Delta_\mathrm{std}$\ as a measure of the clustering of these $k$\ most concentrated orbital poles.

\subsubsection{Pre-Gaia 2018 sample}
Before the Gaia DR2 (Pre-Gaia 2018 sample of proper motions), six of the 11 classical satellite galaxies had their orbital poles clearly associated with a common direction. The orbital poles of the Large and Small Magellanic Clouds (LMC and SMC), Draco, Ursa Minor, Fornax, and Leo\,II all cluster close to the normal vector to the near-polar best-fit plane to the satellite positions (for both the classical satellites only, VPOSclass, as well as including fainter Milky Way satellites, VPOSnew), as well as close to the normal vector to the Magellanic Stream. Their spherical standard distance is only $\Delta_\mathrm{std}(k=6) = 15^\circ.4$\ (those for other $k$\ are compiled in Table \ref{tab:deltasph}).
Sculptor's orbital pole is $180^\circ$\ from this directon, meaning its orbit is aligned with the same plane though it is counter-orbiting with respect to the majority of satellites. In addition, Carina's orbital pole is not well constrained but consistent with this common direction within 1 to 2 $\sigma$. The innermost of the 11 classical satellites, Sagittarius, is clearly unassociated. While its orbit is also polar relative to the Milky Way, is is also nearly perpendicular to the preferred orbital plane of the majority of classical satellites. Finally, Sextans and the most distant of the 11 classical satellites, Leo\,I, both have orbital poles away from the main cluster of orbital poles and closer to the South Galactic Pole, making them somewhat prograde with respect to the Galactic rotation. Interestingly, Leo\,I's 3D velocity vector indicates that it is moving towards and along a planar structure of non-satellite galaxies in the North of the Milky Way, many of which are likely backsplash candidates \citep{2012MNRAS.426.1808T, 2014MNRAS.440..908P}.

\subsubsection{Gaia DR2 Only sample}
For the Gaia DR2 Only sample, the proper motions are entirely independent of the Pre-Gaia 2018 sample. Yet, the overall situation is remarkably similar. Again six orbital poles cluster close in the center of the plot around $(l,b) = (180^\circ, 0^\circ)$: They have $\Delta_\mathrm{std}(k=6) = 16^\circ.8$, which is only $1^\circ.4$\ larger than for the Pre-Gaia 2018 sample of proper motions. These are mostly the same satellites as for the Pre-Gaia 2018 sample, with one notable change: Carina's orbital pole is now considerably better constrained than previously and is clearly part of the cluster. At the same time, the most-likely orbital pole of Leo\,II is about $90^\circ$\ away from the cluster. Due to their large distances, the proper motion uncertainties of Leo\,I and II from Gaia DR2 are quite large, resulting in only weakly constrained orbital poles. Consequently, both are consistent within their $1\,\sigma$\ uncertainties to align with the main cluster of orbital poles. The orbital poles of Sagittarius, Sextans, and Sculptor are approximately identical to those found for the Pre-Gaia 2018 sample of proper motions. The latter thus remains well aligned but counter-orbiting, wheras the former two are clearly not associated with the cluster of orbital poles.

\subsubsection{Combined sample}

\begin{figure*}
	\includegraphics[width=\columnwidth]{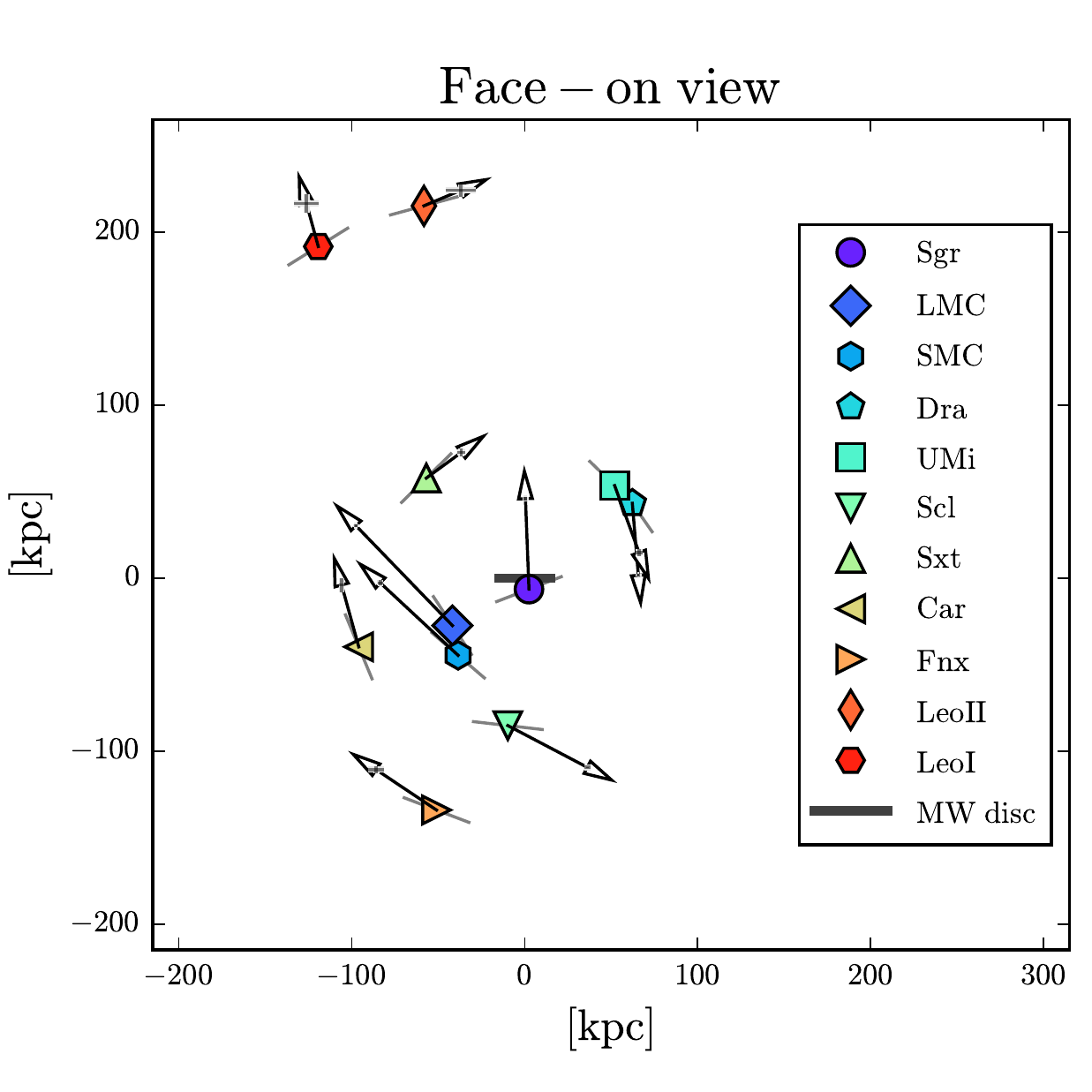}
	\includegraphics[width=\columnwidth]{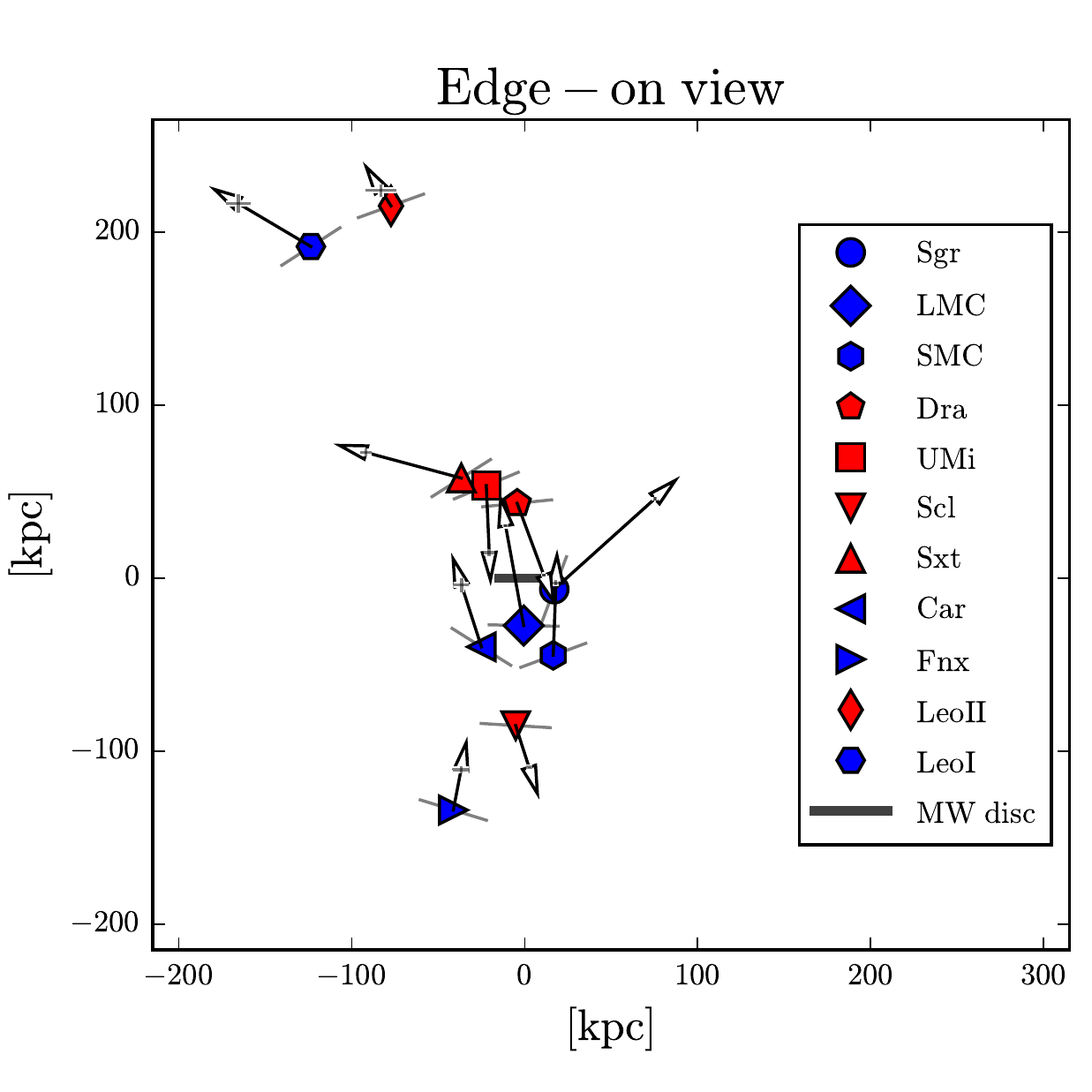}
    \caption{
Most-likely 3D velocity vectors of the 11 classical satellite galaxies (black arrows), projected onto a face-on view (left panel) and an edge-on view (right panel) of the average orbital plane determined from the seven best-aligned orbital poles. The absolute lengths of the vectors is arbitrary, but their relative lengths represent the relative velocities of the satellite galaxies. For both panels, the measurement uncertainties in the two plotted velocity components (along the vertical and horizontal axes) are indicated by the grey error bars at the base of the arrow heads. These use the same scale as the velocity vectors. The error bars thus illustrate how little the vectors are allowed to move around within the uncertainties: The errors are generally much smaller than the arrow heads, except for the most distant Milky Way satellites. The grey lines indicate the tangential direction at each satellite's position. Most satellites have highly tangentially biased velocites in the face-on view, but much more radial velocities in the edge-on view. With the exception of Leo\,I, Sextans, and Sagittarius, the classical satellites of the Milky Way thus move predominantly in a common plane. In the edge-on view (right panel) the satellites are color-coded according to their line-of-sight velocity: red for receding, blue for approaching in this projection. A common line-of-sight velocity trend indicative of rotation is clearly visible: four of five satellites in the north are receding, five of six satellites in the south are approaching. This is similar to the line-of-sight velocity trends observed for the M31 and Centaurus\,A satellite planes.}
    \label{fig:edgefaceon}
\end{figure*}

Unsurprisingly, the Combined sample of proper motion confirms this overall picture. However, in this case seven of the 11 classical satellite galaxies are part of the dominant cluster of orbital poles, with $\Delta_\mathrm{std}(k=7) = 16^\circ.0$.  This is because the Gaia DR2 sample has superior proper motions for Carina which place it among the cluster of orbital poles, whereas for the most distant dSphs Leo\,I and II the HST proper motions of \citet{2013ApJ...768..139S,2016AJ....152..166P} dominate due to their smaller uncertainties, which places Leo\,II among the cluster. The average direction of these orbital poles aligns with the VPOSclass normal vector to $\theta_{\mathrm{VPOSclass}}(k=7) = 22^\circ.1$, and is even closer to the VPOSnew ($\theta_{\mathrm{VPOSnew}}(k=7) = 16^\circ.9$). In addition, Sculptor is counter-orbiting close to the same orbital plane, such that only Sagittarius, Sextans, and Leo\,I can not be clearly associated to the VPOS. For Sagittarius, its position almost perpendicular to the VPOS prohibits a close alignment of its orbital pole with the VPOS for any possible proper motion (the closest possible alignment is approximately $60^\circ$, see \citealt{2013MNRAS.435.2116P}). For Sextans, the orbital pole is well enough constrained by the available proper motions to rule out a close association, even though its position is close to the common VPOS plane and would have allowed a close alignment of its orbital pole. The orbital pole of Leo\,I is at most consistent to 2-3\,$\sigma$\ with the main cluster of orbital poles. 

These findings can also be seen in the current velocity vector directions, as shown in Fig. \ref{fig:edgefaceon}. In an face-on view (left panel) of the preferred orbital plane, a majority of satellite galaxies follows a common velocity trend, clockwise in the plot. For most satellites, the tangential velocity component dominates in this view\footnote{For a discussion of the tangential velocity excess of the classical satellite galaxies relative to $\Lambda$CDM expectations, see \citet{2017MNRAS.468L..41C}}. This is different if the view is rotated by $90^\circ$\ into an edge-on orientation (right panel). In this projection, the satellites that move close to a common plane have predominantly radial velocity components.  The three previously-discussed outliers can also be clearly identified in this view from their considerable velocity components in the horizontal direction of the plot. 
For the edge-on view, the sign of a satellite's velocities perpendicular to the plot is indicated by the color of the symbols. It shows a clear trend, with satellites on one side predominantly receding and those on the opposite side predominantly approaching the viewer. This strongly resembles the coherent line-of-sight velocity correlations observed for the satellite planes of Andromeda \citep{2013Natur.493...62I} and Centaurus\,A \citep{2018Sci...359..534M} that both are observed close to edge-on. If we were to see the Milky Way satellite system in a similar edge-on orientation, we would thus observe a similar line-of-sight velocity trend. Since in the case of the Milky Way we know that the full three-dimentional velocities show that a majority of the classical satellites co-orbits along a common plane, this therefore lends support to the interpretation of the velocity correlations in external systems as indicative of rotating satellite planes.

We also test whether the overall distribution of the orbital poles relative to the direction of the minor axis of the satellite galaxy distribution is consistent with that expected for a random situation, by employing the Kuipers test described in \citet{2008ApJ...680..287M}. This compares the cumulative distribution of the orbital poles from the direction of the minor axis of the satellite galaxy positions with the fitting function for the cumulative biased random pole distance distribution provided by \citet{2008ApJ...680..287M}. We find that for the Combined sample, we can exclude the possibility that the observed cumulative pole distance distribution is drawn from the random one with over 99 per cent significance. For the Pre-Gaia 2018 and the Gaia DR2 only samples, this significance lies at approximately 98 and 95 per cent, respectively.

% Never tell me the odds … ah well, in a paper it might in fact make sense.

In summary, compared to the pre-Gaia situation, the new, independently measured proper motions confirm the previous findings of \citet{2008ApJ...680..287M} and \citet{2013MNRAS.435.2116P} that the majority of the classical satellite galaxies co-orbit close to a common plane. Specifically, the most important additional information provided by the Gaia DR2 proper motions is that they now also place Carina firmly into the group of satellites with aligned orbital poles. Furthermore, the inclusion of the Gaia DR2 data increases the overall clustering of orbital poles for all $k > 5$, both relative to the 6-year old results of \citet{2013MNRAS.435.2116P} as well as relative to the situation just prior to Gaia DR2.

\subsection{Time-Evolution of Orbital Pole Concentration}
\label{subsec:timeevolution}

\begin{figure}
	\includegraphics[width=\columnwidth]{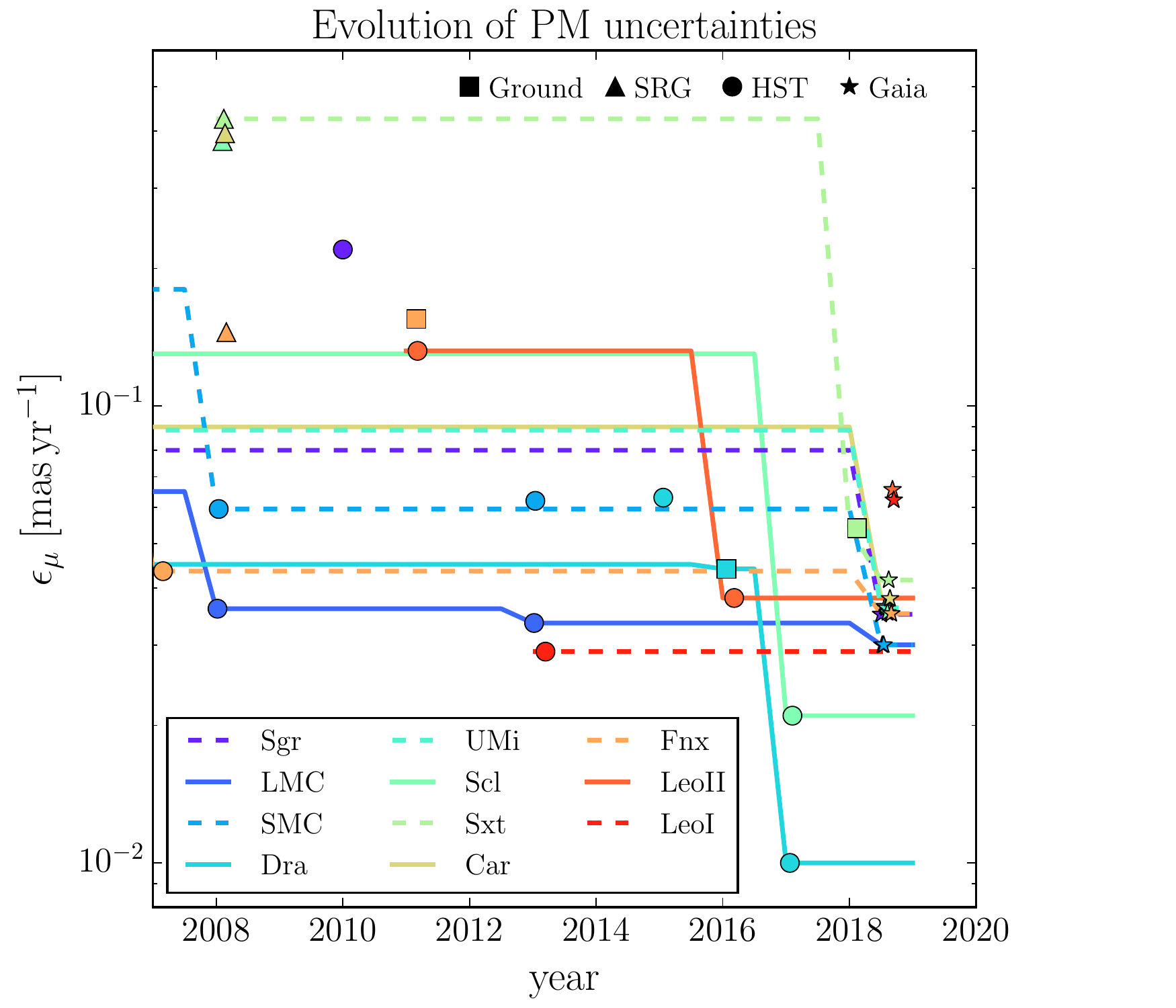}
	\includegraphics[width=\columnwidth]{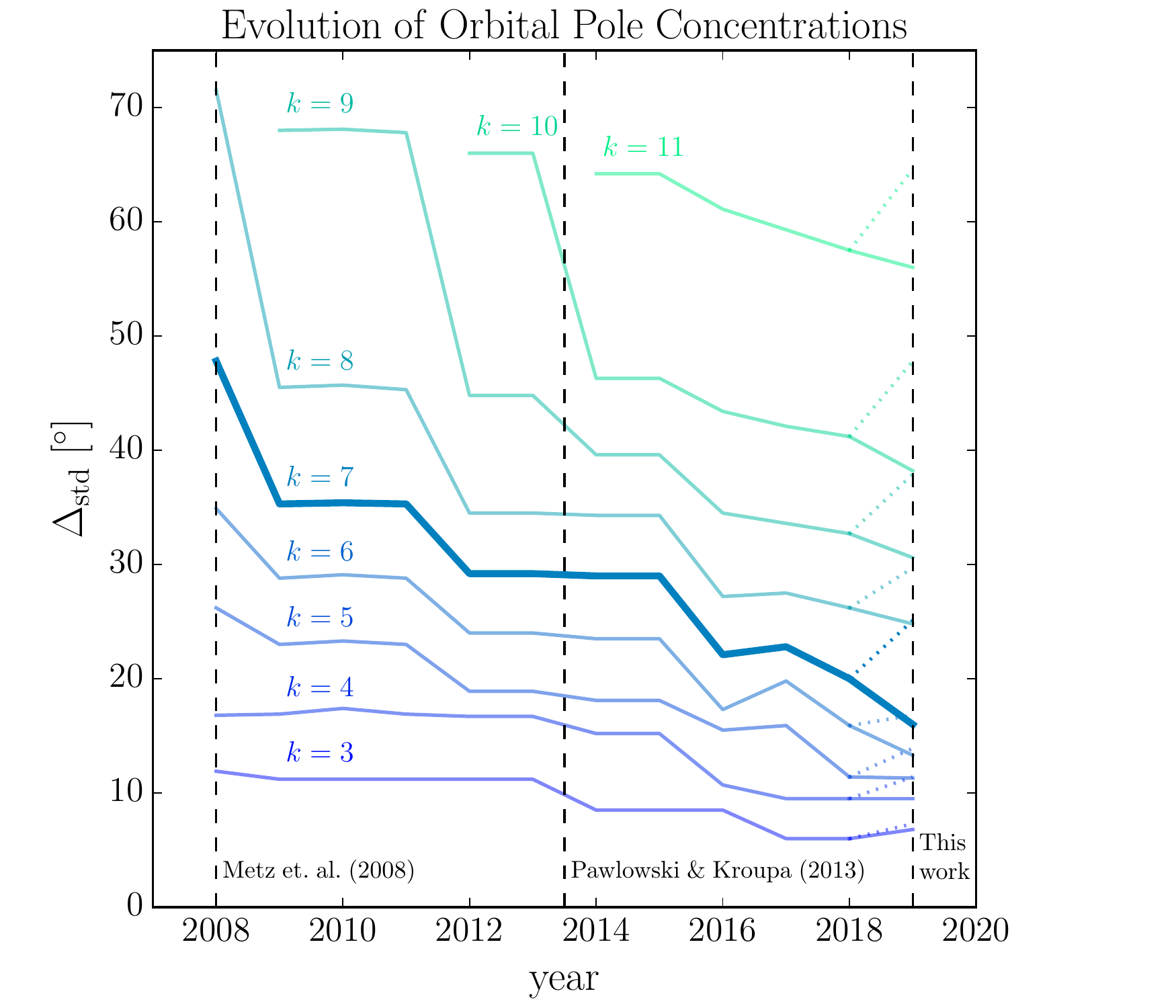}
    \caption{
      Evolution of the smallest reported PM uncertainty $\epsilon_\mu$\ at the time (top panel, lines indicate the smallest reported proper motion uncertainty available in a given year) and the spherical standard distance $\Delta_\mathrm{std}$\ of the $k$\ most-concentrated orbital poles (bottom panel). As new and better measurements became available and the PM uncertainties were reduced, the orbital poles also became successively more concentrated. Vertical dashed lines in the bottom panel indicate the approximate publication times of publications investigating the MW satellite orbital pole distribution: \citet{2008ApJ...680..287M}, \citet{2013MNRAS.435.2116P}, and this work.
    }
    \label{fig:history}
\end{figure}

If an underlying distribution is highly aligned or correlated, then its observational signature should become increasingly present as observational uncertainties decrease. This is because random errors from measurement uncertainties introduce a dispersion, which is larger the larger the uncertainties are \citep[see for example the discussion in][]{2017AN....338..854P}. It is thus interesting to investigate the evolution of the concentration of orbital poles, as measured with $\Delta_\mathrm{std}$, and compare it with the evolution of reported measurement uncertainties of observed proper motions.

The upper panel of Figure \ref{fig:history} shows the reported uncertainties $\epsilon_\mu$\ of proper motion measurements for the 11 classical satellite galaxies for publications between 2007 and 2018, where $\epsilon_\mu$\ is the average of the errors reported for the two proper motion components. The lines indicate the smallest reported proper motion uncertainties available in the published literature at the beginning of a given year\footnote{That means that for example for 2014, all proper motion measurements with a publication date of 2013 or earlier are considered to be available, and we only consider published works but no preprints or private communications.}. There is a clear trend to smaller $\epsilon_\mu$ with time, and for several of the classical satellites the Gaia DR2 proper motions (star symbols) have resulted in a substantial decrease in $\epsilon_\mu$.

The lower panel of Figure \ref{fig:history} shows the evolution of the spherical standard distance $\Delta_\mathrm{std}$\ as better proper motion data became available. For each year, the best available proper motion measurements of a given technique (ground, stellar redshift gradient, Hubble Space Telescope) that was published until the end of the previous year is chosen, and the uncertainty-weighted average of these different techniques calculated. The resulting average proper motion is then used to construct the orbital poles, and the spherical standard distance of all $k$\ best-aligned poles is calculated. For years before 2014 not all of the 11 classical satellites had proper motion measurements available, such that the maximum $k$\ is smaller than 11. The vertical lines indicate the situation present at the approximate time of two previous publications, \citet{2008ApJ...680..287M} and \citet{2013MNRAS.435.2116P}. 

As the proper motion measurements become more precise, $\Delta_\mathrm{std}$\ becomes successively smaller for all $k$. Almost all lines show a clear, monotonic decrease. This is not merely driven by the addition of more satellites to the sample. Since 2013, there has been at least one reported proper motion measurement for each of the 11 classical satellites. Yet $\Delta_\mathrm{std}$\ has since continued to decrease for every $k$, not only for the present comparison but also compared to the $\Delta_\mathrm{std}$\ values reported by \citet{2013MNRAS.435.2116P} as listed in Table \ref{tab:deltasph}. This trend continues with the inclusion of Gaia DR2 proper motions (last step), with the exception of the $k=3$\ line which shows a minor increase.

The dotted lines in the lower panel of Fig. \ref{fig:history} indicate what would happen if only Gaia DR2 proper motion measurements would be considered from 2018, and all previous measurements would be dismissed (including those with superior accuracy, most notably the HST proper motions of Leo\,I and II). Compared to the best available pre-Gaia DR2 proper motion measurements at the end of 2017, $\Delta_\mathrm{std}$ would increase somewhat for all $k$. However, the deduced clustering of orbital poles would still be tighter than, or comparable to, that found in 2013. Thus, the Gaia DR2 proper motions of the 11 classical satellites clearly support and thus reinforce the findings reported in \citet{2013MNRAS.435.2116P}, in contrast to the qualitative discussion in \citet{2018A&A...616A..12G} which could be interpreted differently. The evolution of the orbital pole concentration with increasing measurement accuracy is thus consistent with that expected for a correlated distribution: more precise measurements find a tighter clustering of orbital poles. The inclusion of Gaia DR2 proper motions of the 11 classical satellite galaxies confirms this tendency.

\section{Discussion}
\label{sec:Discussion}

\subsection{Expected Orbital Pole Concentration for Random Velocities}

\begin{figure}
	\includegraphics[width=\columnwidth]{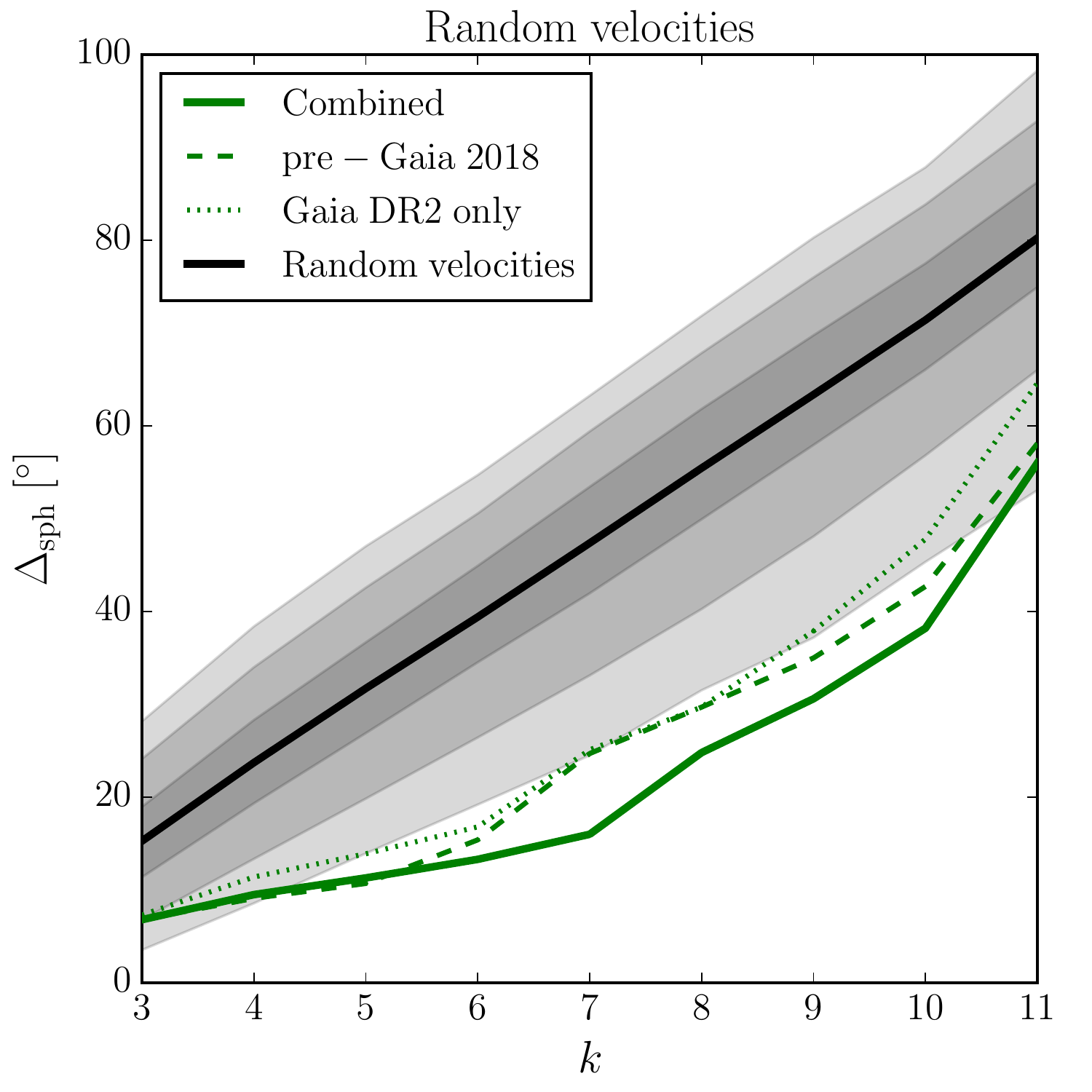}
    \caption{
      Comparison of the spherical standard distances $\Delta_\mathrm{std}$\ of the $k$\ most-concentrated orbital poles with the expected distribution from 2500 realizations of random velocity vectors drawn from isotropy of the classical satellites (but fixed observed 3D positions). The three observational proper motion samples are shown as green lines. Contours on the model distribution indicate regions containing 50, 90 and 99 per cent of all realizations.
    }
    \label{fig:concen_random}
\end{figure}

As a measure of the significance of the orbital pole alignment, we now determine how frequently similarly small $\Delta_\mathrm{std}$\ occur if the satellites had random velocities. For that, we adopt the observed positions of the satellites, but chose the direction of their 3D velocity vector from an isotropic distribution. Thus, each satellite's orbital pole will be confined to the observationally allowed great circle defined by its current position vector relative to the Milky Way. For each proper motion sample and each number $k$ of combined orbital poles we then calculate the frequency $f_\mathrm{random}$\ within $N = 2500$\ such random realizations that have an as small or smaller $\Delta_\mathrm{std} (k)$. Here $N$\ was chosen to be comparable to the number of systems identified in the Illustris TNG100-1 cosmological simulation (see Sect. \ref{sec:illustris}).

The results are compiled in Table \ref{tab:deltasph}, and illustrated in Fig. \ref{fig:concen_random}. With the inclusion of more recent proper motion measurements, the significance of the orbital pole alignments has increased relative to \citet{2013MNRAS.435.2116P}. Even considering the Gaia DR2 sample only, the frequencies $f_\mathrm{random}$\ are very comparable to those for the 2013 proper motions, and only slightly larger than the most recent pre-Gaia proper motions. The Gaia DR2 proper motions thus can not be said to reduce the significance of the kinematic coherence of the 11 classical satellite galaxies in the VPOS. Instead, considering them as an independent measurement, they confirm the overall findings based on other proper motion techniques: the orbital poles of the Milky Way satellites are substantially clustered. The full strength of the new data comes to play when combining it with the most accurate previous proper motions in the Combined sample. This reveals an even tighter alignment of orbital poles and consequently very high significance, for $k = 7$\ and 8 in fact none of the 2500 random realizations reproduces a similarly tight clustering as observed (significance $\geq 99.96$\,per cent), while for $k = 5$\ to 10 the significance exceedes 99.9\ per cent ($f_\mathrm{random} < 0.1$\,per cent).

% Significance of 99.9 per cent … So certain, are you.

\subsection{The Best-Possibly-Aligned Orbital Poles}

\begin{figure*}
	\includegraphics[width=140mm]{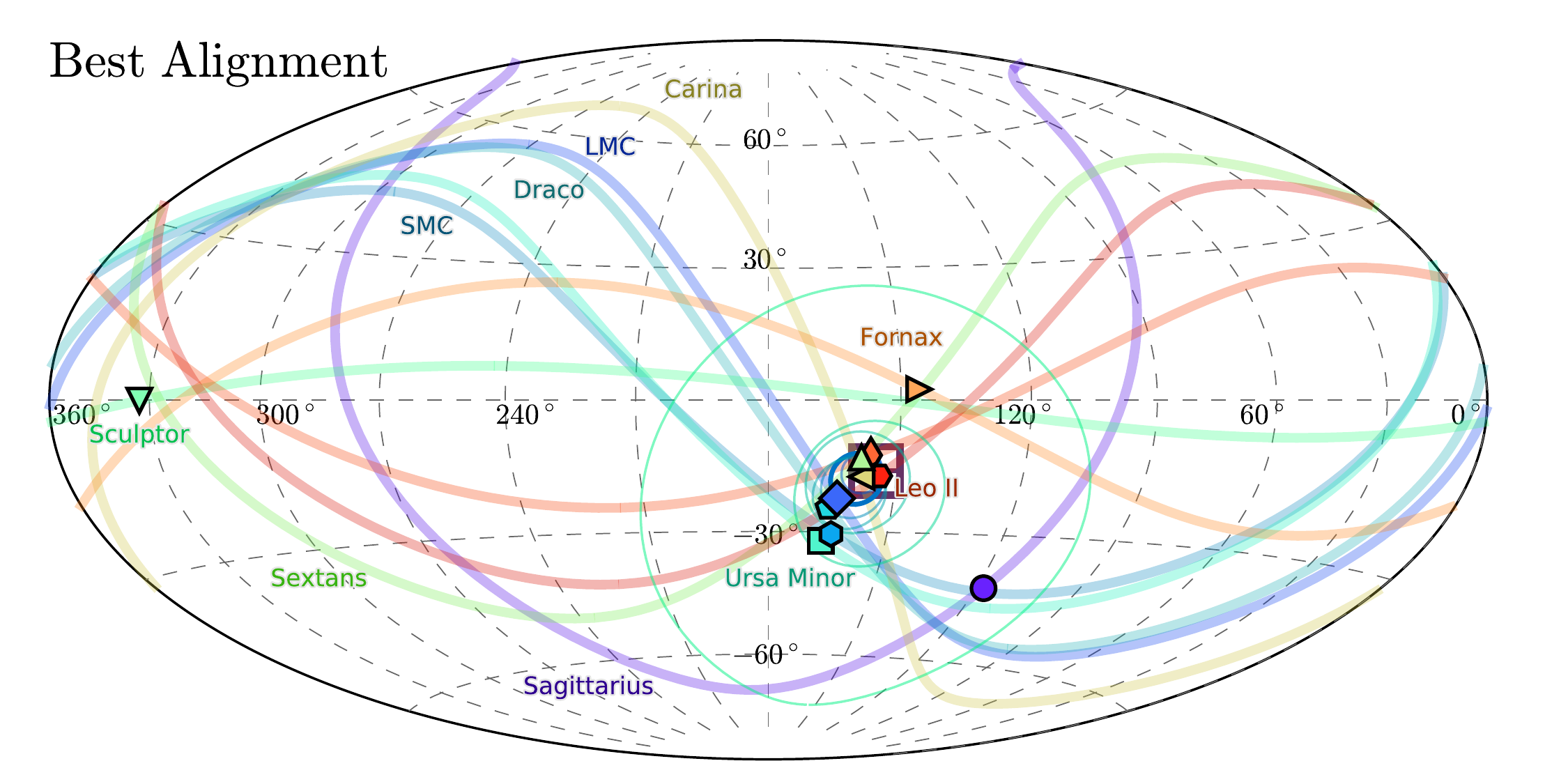}
    \caption{
      Closest possible alignment of orbital poles with the VPOS normal direction (magenta square) given each satellite's current position, independent of measured PM. Sculptor's orbital pole is placed in the counter-orbiting direction. The color-coded great circles indicate all possible orbital pole directions for the 11 classical MW satellites. These are constrained by the requirement that an orbital pole must by definition be perpendicular to the current position of a satellite galaxy. The circles indicate the average direction of the $k = 3\ \mathrm{to}\ 11$\ most concentrated orbital poles, as in Fig. \ref{fig:asps}.
    }
    \label{fig:aspbest}
\end{figure*}

\begin{figure*}
	\includegraphics[width=\columnwidth]{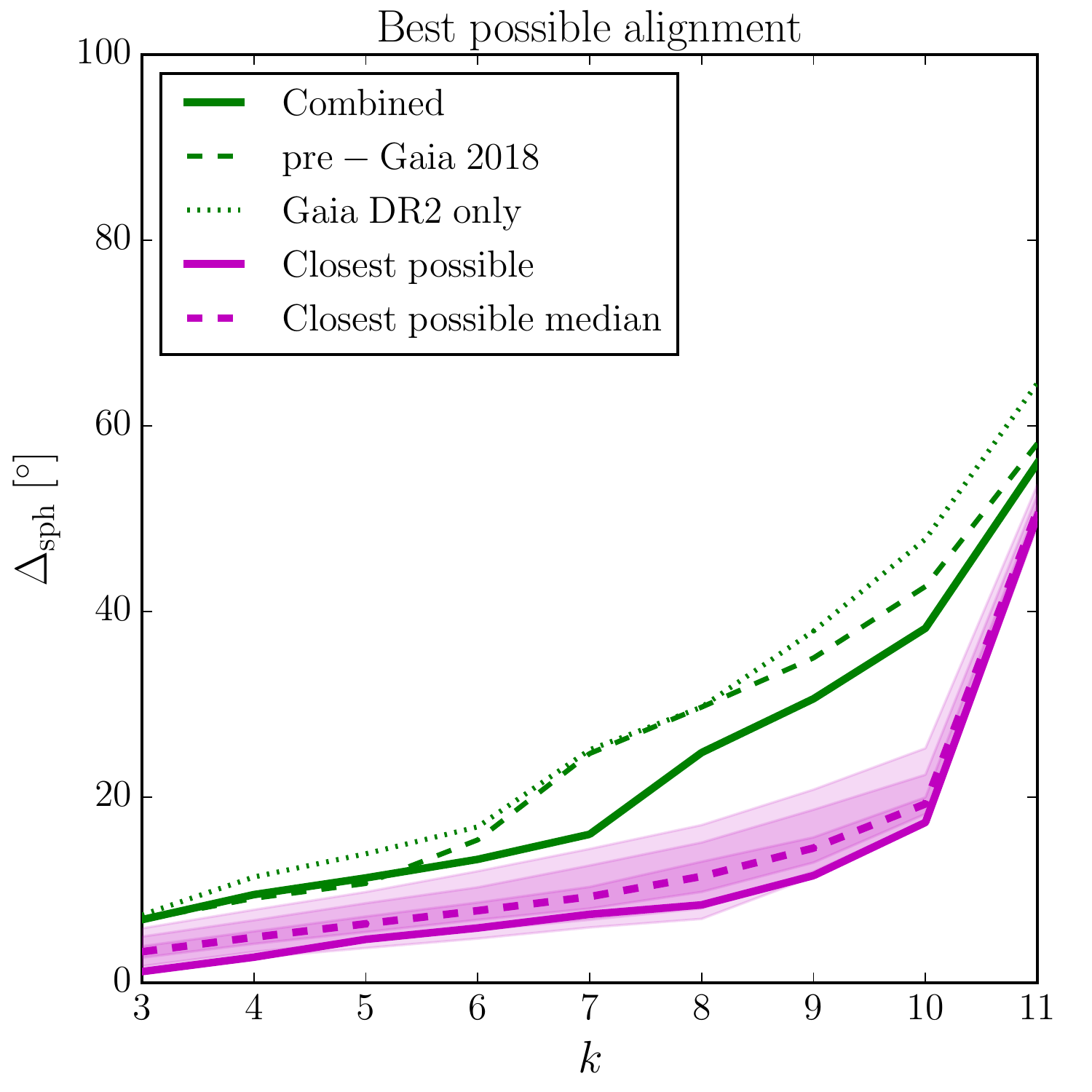}
	\includegraphics[width=\columnwidth]{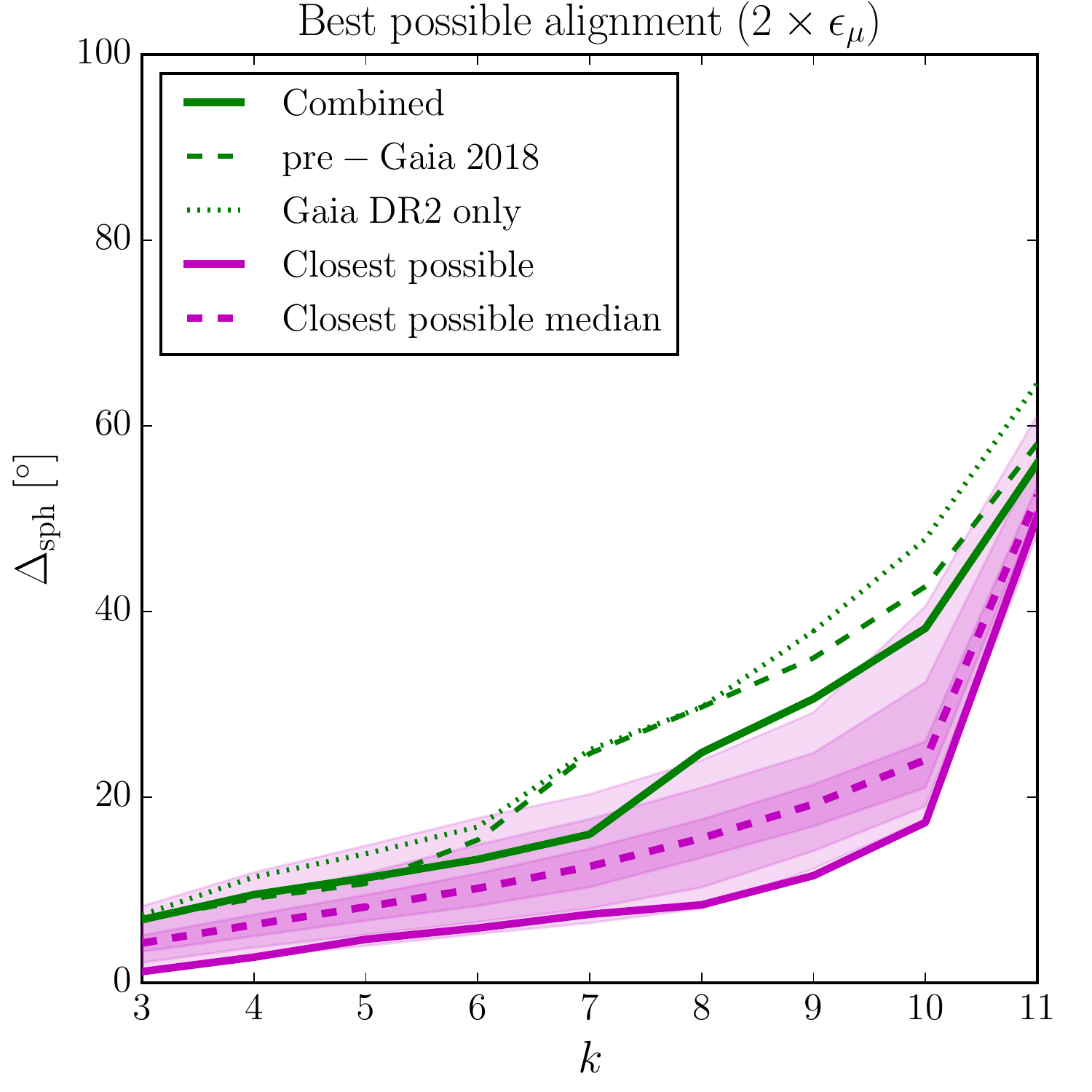}
    \caption{
    Same as Fig. \ref{fig:concen_random}, but comparing to the orbital pole directions that best align with the normal to the VPOSclass. The solid magenta line gives the standard distances for the best-aligned poles, while the dashed line and the shaded regions are derived from 2500 realizations of mock-observing this while accounting for proper motion uncertainties of $\epsilon_\mu = 0.05\,\mathrm{mas\,yr}^{-1}$\ (left panel) and $\epsilon_\mu = 0.1\,\mathrm{mas\,yr}^{-1}$\ (right panel).
    }
    \label{fig:concen_best}
\end{figure*}

If the Milky Way satellites were aligned in a perfect plane containing the center of the Milky Way, then they could all share the exact same orbital pole (which would be identical to the normal vector of the perfect plane), and $\Delta_\mathrm{std} (k) = 0^\circ$\ would in principle be achivable. However, the Milky Way satellites are not in a perfect planar arrangement. The observed distribution of the 11 classical satellites has a short-to-long axis ratio of $c/a = 0.182$, or a physical root-mean-square height of $r_\mathrm{per} = 19.6\,\mathrm{kpc}$\ \citep[e.g.][]{2014ApJ...789L..24P}. Thus, they can not all share the exact same orbital plane, and consequently $\Delta_\mathrm{std} (k) > 0^\circ$\ (for $k>2$). However, for a given plane passing through the center of the Milky Way, each satellite has one possible orbital pole direction that comes closest to the normal vector chosen to define the plane.

To study the extent to which the spatial thickness of the observed satellite distribution constrains the minimum achivable scatter in $\Delta_\mathrm{std} (k)$, we determine the best-possibly aligned orbital poles for the VPOSclass plane, i.e. the plane minimizing the distances to the 11 classical satellite galaxies. For each satellite, we adopt its observed position and assign it a velocity direction that is perpendicular to both the position vector and the VPOSclass normal vector. Implicitly this assumes that each of the satellites is right now at its furthest perpendicular offset it ever reaches from their common plane during its orbit.
Furthermore, to illustrate the substantial effect a single counter-orbiting satellite has on  $\Delta_\mathrm{std} (k=11)$, we require Sculptor's orbital pole to be opposite to the others. The distribution of best-aligned orbital poles thus determined is shown in Fig. \ref{fig:aspbest}.

In addition to these closest-possibly-aligned orbital pole directions, we also generate 2500 samples accounting for the uncertainty in proper motion measurements. If a distribution is highly clustered, such uncertainties will always tend to add additional dispersion and thus reduce the clustering (increase $\Delta_\mathrm{std}$). This is the more realistic \textit{measurable} minimum clustering expected if all satellites were co-orbiting close to the same plane. For this, we assign each satellite with a tangential velocity of $175\,\mathrm{km\,s}^{-1}$\ (the median tangential velocity for the 11 classical satellites as derived for our combined sample of proper motions), and mock-observed it 2500 times at its actual distance while adding proper motion errors by drawing from a normal distribution with width of $\sigma = 0.05\,\mathrm{mas\,yr^{-1}}$.

The frequency $f_\mathrm{bestaligned}$\ of these 2500 samples that have $\Delta_\mathrm{std}(k)$\ {\it larger} than the observed value is compiled in Table \ref{tab:deltasph} for the Combined sample. It is largely zero for the other samples. For $k=7$ combined orbital poles, about 0.5 per cent of these artificially generated satellite velocities result in orbital pole distributions that are in fact wider than the observed one. 
Figure \ref{fig:concen_best} compares the resulting spherical standard distances of the best-possibly-aligned orbital poles with the measured distribution. What is interesting is that the observed $\Delta_\mathrm{std}$\ distribution closely traces the best-possible until $k=7$\ for the Combined sample of proper motions. Up to $k=7$, the slope of $\Delta_\mathrm{sph}(k)$\ is the same as that of the best-aligned orbital poles model, while beyond $k=8$ it is more similar to that with random velocities. This is another strong indication that seven of the eleven classical dwarfs are most consistent with co-orbiting along the common plane as part of a common dynamical structure. 

It is worth noting that slightly increasing the synthetic proper motion error to $\sigma = 0.10\,\mathrm{mas\,yr^{-1}}$\ brings the observed and the typically expected $\Delta_\mathrm{std}$ in close agreement (right panel of Fig. \ref{fig:concen_best}). The distribution of orbital poles of the 11 classical MW satellite galaxies is therefore consistent with the best possible alignment of orbital planes if the reported observed proper motions are affected by unaccounted for errors on the order of $0.1\,\mathrm{mas}$, or if some velocity dispersion perpendicular to the satellite plane of order $40\,\mathrm{km\,s}^{-1}$\ (corresponding to $0.1\,\mathrm{mas\,yr}^{-1}$\ for the median distance to the classical satellites) is present, as to be expected for a structure of non-finite thickness.

It is also clear that having only one counter-orbiting satellite dominates the $\Delta_\mathrm{std}$\ signal at $k=11$.  This can be understood from the definition in Equation \ref{eq:Delta_std}. Assuming that for $k=11$, 10 orbital poles point into the exact same direction while only one points in the opposite direction. Then the latter is $\pi = 180^\circ$\ off from the average direction, while the other 10 have $0^\circ$ offset. The sum in Equation \ref{eq:Delta_std} thus becomes $1 \times \pi^2 + 10 \times 0^2 = \pi^2$, or $\left(180^\circ\right)^2$\ and $\Delta_\mathrm{std}(k=11) = \sqrt{\frac{\pi^2}{11}} = 0.947$, or $\Delta_\mathrm{std}(k=11) = 54.3^\circ$. This is very close to the observed value for the combined proper motion sample, which is $\Delta_\mathrm{std}(k=11) = 56.0^\circ$. 
This illustrates that the measure $\Delta_\mathrm{std}$\ is not optimal to account for counter-orbiting satellites within a common structure. Consequently, the choice of \citet{2015MNRAS.452.3838C} to consider $\Delta_\mathrm{std}$\ of all 11 classical satellites in their analysis of the look-elsewhere effect biases their results to include many less extreme systems than the actually observed one. This results in a too high frequency of simulated satellite systems that are identified to resemble the Milky Way, in part explaining why they report a higher frequency of simulated systems that are as rare as the observed one (11 per cent) compared to studies using different approaches \citep[e.g.][]{2014MNRAS.442.2362P,2019MNRAS.tmp.1692S}.

\subsection{Specific angular momenta}

\begin{figure}
	\includegraphics[width=\columnwidth]{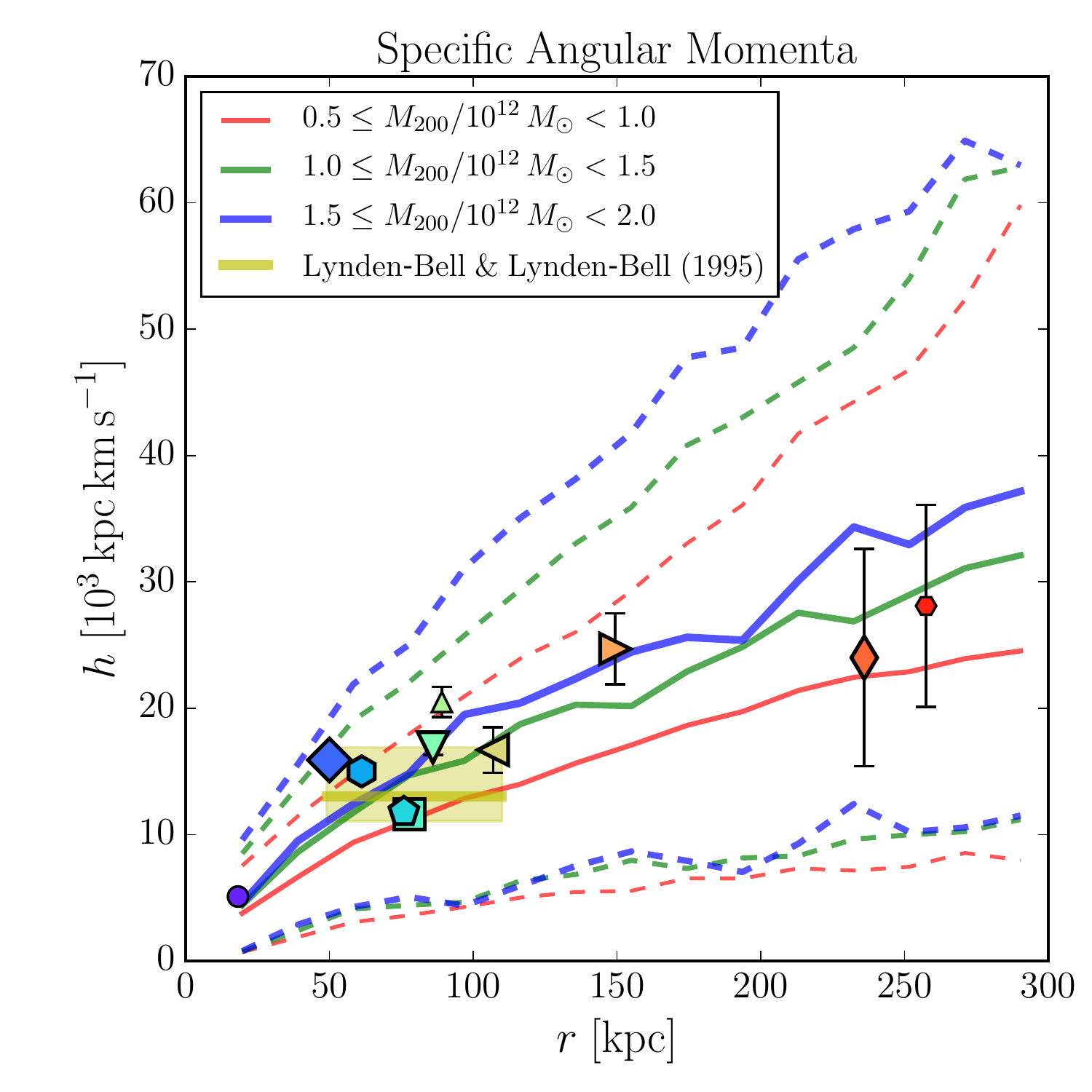}
    \caption{
      Specific angular momenta $h$\ of the 11 classical satellite galaxies, based on the Combined sample of proper motions, plotted against their Galactocentric distances $r$. The markers identify the satellites as in Fig. \ref{fig:edgefaceon}, those satellites that have orbital planes aligned to within $20^\circ$\ of a common plane are highlighted as larger symbols. The yellow line indicates the specific angular momentum \citet{1995MNRAS.275..429L} predicted for six satellite galaxies identified as members of a possible stream; these six are contained in the yellow shaded region. 
      Also shown are the median distributions of specific angular momenta of simulated satellites for hosts of three mass bins (solid lines, with region between dashed lines containing 90 per cent of all simulated satellites). These were obtained from the systems in the Illustris TNG100-1 simulation selected in Sect. \ref{sec:illustris}.
    }
    \label{fig:specangmom}
\end{figure}

In addition to the directions of angular momenta of the satellite galaxies (the orbital poles), we can also compare their specific angular momenta $h$\ (the mass-normalized absolute values of the angular momenta). These are compiled in Table \ref{tab:individualpoles}, and plotted in Fig. \ref{fig:specangmom} against the distance of the satellite galaxies from the Galactic center.

Based on their spatial alignment, distances, and line-of-sight velocities, \citet{1995MNRAS.275..429L} identified the LMC, SMC, Draco and Ursa Minor, as well as possibly Sculptor and Carina, as a potential stream of satellite galaxies. The stream's predicted orbital pole, in part motivated by then-available proper motion measurements of the LMC,  points to $(l, b) = (190^\circ, 3^\circ)$, and the specific angular momentum that the stream members are expected to share if they are indeed physically associated was predicted to be $h = 1.3 \times 10^4\,\mathrm{kpc\,km\,s}^{-1}$\ (shown as a yellow line in Fig. \ref{fig:specangmom}). It is striking how well this prediction is matched given the data available now, 24 years later. Five of the six proposed members orbit in the same direction (the exception being Sculptor), and all six orbit close to the common orbital plane predicted; their orbital poles align to $30^\circ$\ or less. Furthermore, all of the six possible members identified then have specific angular momenta within 30 per cent from the predicted value, the largest difference is with Sculptor (the only counter-orbiting one), and Carina. Intriguingly, these two were also the two possible members \citet{1995MNRAS.275..429L} considered as less likely to be associated.

Interestingly, also the orbital pole of Fornax aligns to within $23.1^\circ$\ with the predicted direction, but its specific angular momentum of $h = 2.47 \times 10^4\,\mathrm{kpc\,km\,s}^{-1}$\ is off by almost a factor of two from the predicted value. This similarity in orbital direction, coupled with their close position to the LMC in projection on the sky, has nevertheless led to recent suggestions that out of the classical satellites, not only the SMC but also both Carina and Fornax were once satellites of the LMC \citep{2019arXiv190401028P,2019arXiv190702979J}.
However, as discussed above, objects which were accreted in a common event, such as a massive dwarf galaxy with its satellites, in addition to orbiting in a common plane are expected to share similar specific angular momenta\footnote{Note here that the concept of accreted satellites (and globular clusters) is physical only in the presence of dark matter due to the need for Chandrasekar dynamical friction to dissipate orbital energy \citep{2015CaJPh..93..169K}.}. Indeed, the LMC and SMC have almost identical specific angular momenta: $h = (1.59 \pm 0.02) \times 10^4\,\mathrm{kpc\,km\,s}^{-1}$\ and $h = (1.50 \pm 0.50) \times 10^4\,\mathrm{kpc\,km\,s}^{-1}$, respectively. Carina has $h = (1.67 \pm 0.18) \times 10^4\,\mathrm{kpc\,km\,s}^{-1}$, which is consistent with the LMC/SMC within its uncertainty. However, Fornax has $h = (2.47 \pm  0.28) \times 10^4\,\mathrm{kpc\,km\,s}^{-1}$. In addition, Fornax, at a galactocentric distance of 150\,kpc, is about three times further from the Milky Way than the LMC. Consequently, the association of Fornax as a satellite of the LMC was disputed by \citet{2017MNRAS.465.1879S}, since its current large distance from the LMC and its radial velocity does not match that of their modelled LMC debris. \citet{ 2019arXiv190709484E} have instead calculated possible past orbits of the observed satellites based on their measured proper motions. They found it very unlikely that Carina was an LMC satellite in the past for the full range of their considered LMC masses, while for Fornax they report a non-negligible probability that it was once bound to the LMC only if the latter has a very high mass of $2.5 \times 10^{11}\,M_{\sun}$.

The close alignment of orbital poles of those classical satellite galaxies orbiting in the VPOS, especially since it is also present for satellites on the opposite side of the Milky Way (Draco, Carina) and at large distances (Fornax, Leo\,I), should thus caution against interpreting similarities in the orbital direction of a satellite and the LMC as conclusive evidence for a past association. There is also a more fundamental reason to be cautious: the need to avoid circular reasoning. We know that the observational situation is such that many satellite galaxies are distributed and orbit in a common structure. To explain this, it has been suggested that satellites might have been accreted as part of a group, or as satellites of satellites (\citealt{2008ApJ...686L..61D,2008MNRAS.385.1365L}, but see \citealt{2009ApJ...697..269M}). If this group infall idea is now tested by comparing orbital directions of satellites, and a detected similarity in orbital directions were to be taken as conclusive evidence for a past association as an LMC satellite, this would constitute circular reasoning. After all, the orbital alignment was the observational feature to be explained by the scenario in the first place.

It is therefore important to find independent confirmation of the proposed satellite association, for example with careful energy and angular momentum considerations. This is all the more important since the cosmological arguments in favor of having a Fornax-like satellite of the LMC are not conclusive either. Both in \citet{2019arXiv190401028P} and \citet{2019arXiv190702979J}, the satellite luminosity function shows sufficient scatter that as little as zero satellites with stellar mass $M_{*} > 10^7\,M_{\sun}$, the mass scale encompassing both Fornax and the SMC, can be expected in the simulations. Furthermore, \citet{2019arXiv190401028P} consider as satellites of their LMC analogues everything within two virial radii. By definition, the virial radius $r_{200}$\ scales with virial mass $M_{200}$\ as $r_{200} \propto M_{200}^{1/3}$. The LMC has about one eigth of the virial mass of the Milky Way ($M_{200}^\mathrm{LMC} \approx \frac{1}{8} M_{200}^\mathrm{MW}$). Thus, the volume from which they select their LMC satellites is comparable to the full virial volume of the Milky Way. Consequently, in this picture LMC satellites could be situated anywhere in the Milky Way halo, and the large extent of the considered cloud of LMC satellite systems itself prohibits identification of LMC-association based soley on orbital alignment. The reason is that such LMC satellites would not have to share the same orbital plane around the Milky Way, because such a large extent of a potential satellite cloud makes it possible that a satellite passes the center of the Milky Way on a different side than the LMC itself, resulting in a very different direction of orbital pole relative to the Galaxy.

\section{Comparison to $\Lambda$CDM simulations}
\label{sec:illustris}

% Comparing to LCDM simulations? It's a trap!

\citet{2009MNRAS.399..550L} determined the fraction of systems in the High-Resolution Counterpart Millennium Simulation \citep{2008MNRAS.387..536G} for which $N_\mathrm{satellites}$\ out of 11 satellite galaxies have orbital poles that align with the average direction $J^6_\mathrm{mean}$\ of angular momentum of the six most clustered orbital poles. Specifically, their figure 9 shows that the fraction of simulated systems drops to zero for four orbital poles aligned to within $15^\circ$, or seven orbital poles aligned to within $30^\circ$. The corresponding measure $\theta_\mathrm{J6}$\ of the angle between $J^6_\mathrm{mean}$\ and the individual orbital poles of observed MW satellites is listed in Table \ref{tab:individualpoles}. This shows that four satellites align to better than $12^\circ$\ with $J^6_\mathrm{mean}$\ (LMC, Leo\,I, Draco, and the SMC), and another three have $\theta_\mathrm{J6} < 30^\circ$\ (Ursa Minor, Carina, and Fornax). Thus, the observed alignment of orbital poles is inconsistent with these $\Lambda$CDM predictions.

Cosmological simulations have advanced considerably over the past decade, most notably by increasing the resolution, covered volume, and considered physical effects. To provide a fair comparison with the most up to date expectations, we therefore also compare the alignment and kinematic coherence of the 11 classical satellites to a state-of-the-art hydrodynamical cosmological simulation. In the context of the present study we are particularly interested in determining whether the inclusion of Gaia DR2 proper motions increases the tension between the observed clustering of orbital poles of the classical Milky Way satellites compared to previous proper motion data, and if so by how much. A more detailed analysis of the properties of the best-matching systems in the simulation is left for future work since it goes beyond the scope of this paper and its focus on the observed satellite system.

We use the IllustrisTNG project, specifically the hydrodynamical simulation TNG100-1 and its dark-matter-only equivalent TNG100-1-Dark \citep{2018MNRAS.477.1206N,2018MNRAS.475..676S,2018MNRAS.480.5113M,2018MNRAS.475..624N,2018MNRAS.475..648P}. The simulation has a box size of $75 \mathrm{Mpc}/h$\ at $z = 0$\ and a dark matter particle mass of $m_\mathrm{DM} = 7.5 \times 10^6\,M_{\sun}$\ ($m_\mathrm{DM} = 8.9 \times 10^6\,M_{\sun}$\ for TNG100-1-Dark). This provides a good compromise between simulation volume (to ensure a sufficiently large sample of potential host galaxies with masses comparable to that of the Milky Way), and resolution (so that most of the selected hosts in the Milky Way mass range are indeed surrounded by at least 11 subhalos). The adopted cosmological parameters of Illustris TNG are consistent with \citet{2016A&A...594A..13P}. We use the publicly available redshift zero galaxy catalogs \citep{2019ComAC...6....2N}.

\subsection{Selection and analysis of simulated satellite systems}

As potential Milky Way analogues, we first select all halos with a total mass $M_{200}$\ of 0.5 to $2.0 \times 10^{12}\,M_{\sun}$\ within a sphere containing a mean density of 200 times the critical density of the universe at $z = 0$.  There are 2660 halos in this mass range in TNG100-1. Furthermore, to ensure a sufficient isolation, we require each potential host halo to not have another halo more massive than  $M_{200} = 0.5 \times 10^{12}\,M_{\sun}$\ within 600\,kpc. This distance was chosen to be comparable to the current separation between the Milky Way and M31 of $\approx 790$\,kpc, and to ensure that the volumes within which satellites are considered do not overlap for neighboring hosts. This leaves us with 2646 potential hosts.

For each host, we generate a list of all subhalos between 15 and 300\,kpc from its center. The list begins with subhalos that have a luminous component, ranked by their $V$-band luminosity such that the brightest leads the list. This is then followed by all dark subhalos (no luminous component), ranked according to their dark matter mass. For each host, the top 11 subhalos from this list are then selected as the simulated satellite system. We discard all systems which contain a sub-halo more massive than 20 per cent of the host to avoid selecting systems in the process of a major merger. Of the 2646 potential hosts, 2548 have a sufficient number of subhalos present and are thus part of our further analysis. 

If we had required each subhalo to have a luminous component, only 185 hosts would have contained at least 11 luminous satellites. This would not provide us with a sufficient sample size to measure the frequency of systems as extreme as the observed VPOS. It would also prohibit a fair comparison to the dark-matter-only equivalent TNG100-1-Dark, due to the stochasticity of populating low-mass dark matter halos with a luminous component that could be represented by as few as one stellar particle. We did check and confirm that the distributions of the satellite plane parameters (flattening, kinematic coherence) agree between the full sample of simulated systems and that containing only the subset for which each of the 11 sub-halos has a luminous component.

In selecting subhalos, it is important to exclude those not of cosmological origin (SubhaloFlag = 0 in the IllustrisTNG database). These can be baryonic fragments in the host galaxy's disk instead of an independent satellite galaxy, and would thus rather trace the flattening and rotation of the host galaxy itself which would artificially increase the number of systems with flattened spatial distributions and coherent motions.

To investigate the effect the inclusion of baryonic physics has on the results, we repeat the same analysis for the Illustris TNG100-1-Dark run, the dark-matter-only equivalent of TNG100-1. For this, we find 2628 halos with  $M_{200} = 0.5\ \mathrm{to}\ 2.0 \times 10^{12}\,M_{\sun}$, of which 2616 fulfill the isolation criterion. Of these potential hosts, 2564 do not have a satellite more massive than 20 per cent of the host and contain at least 11 subhalos beyond 15 and within 300\,kpc and are thus considered for our analysis.

For these selected sub-halo satellite systems, we then determine the spatial flattening using a standard tensor of intertia technique \citep{2015ApJ...815...19P}. The flattening of the distribution of 11 satellite sub-halos is measured both in a relative sense via the root-mean-square minor-to-major axis ratio $c/a$, as well as via the absolute root-mean-square height $r_\mathrm{per}$\ perpendicular to the plane minimizing the sub-halo distances (i.e. along the minor axis). For the classical satellites of the Milky Way, the observed flattening is $c/a = 0.182$\ and $r_\mathrm{per} = 19.6\,\mathrm{kpc}$\ \citep[e.g.][]{2014MNRAS.442.2362P,2014ApJ...789L..24P}. In addition, we calculate each sub-halo's orbital pole and measure the concentration of the $k = [3, ..., 11]$\ most concentrated ones following the method described in Sect. \ref{sec:OrbPoles}. We also determine the intermediate-to-major axis ratio $b/a$.

\subsection{Results}

\begin{figure*}
	\includegraphics[width=\columnwidth]{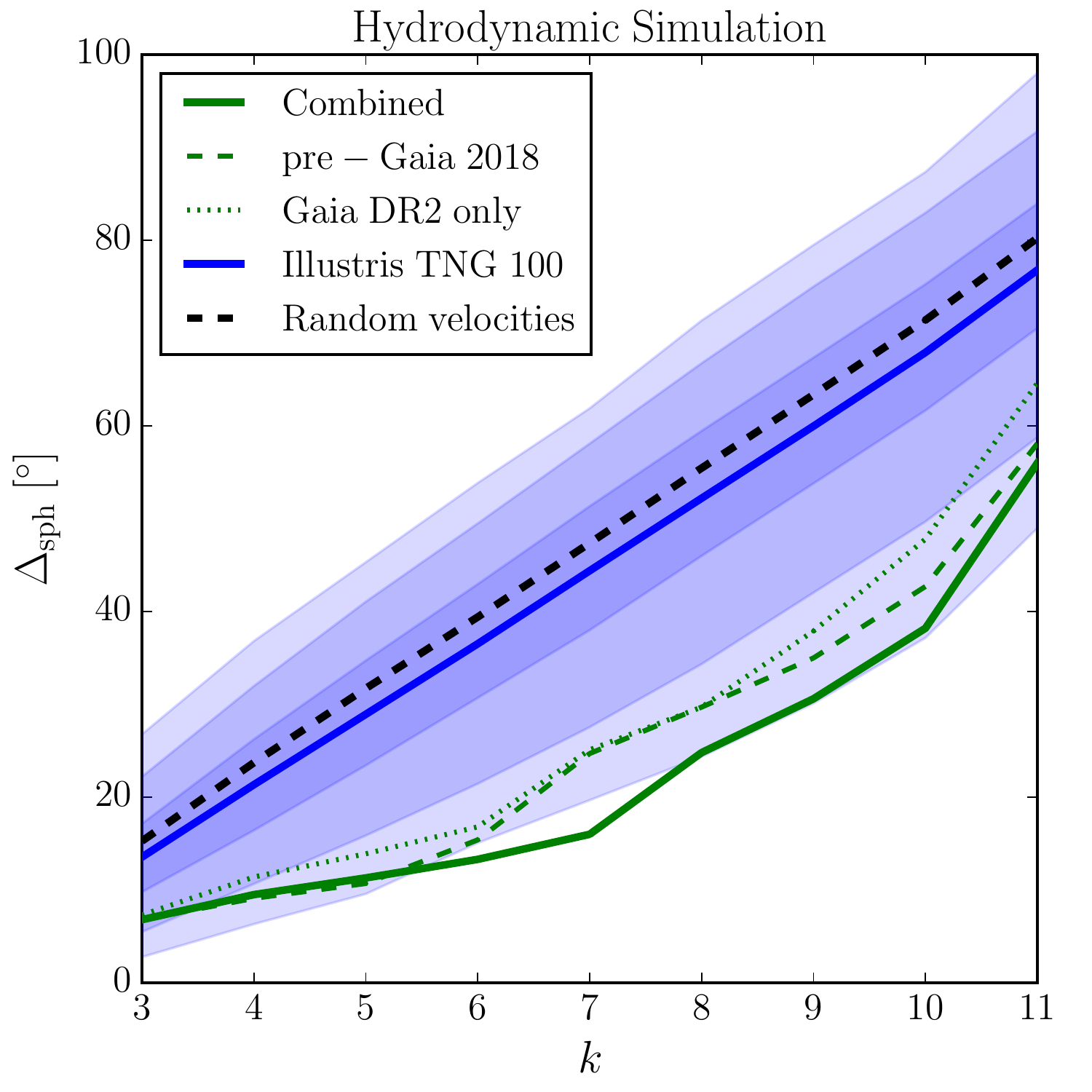}
	\includegraphics[width=\columnwidth]{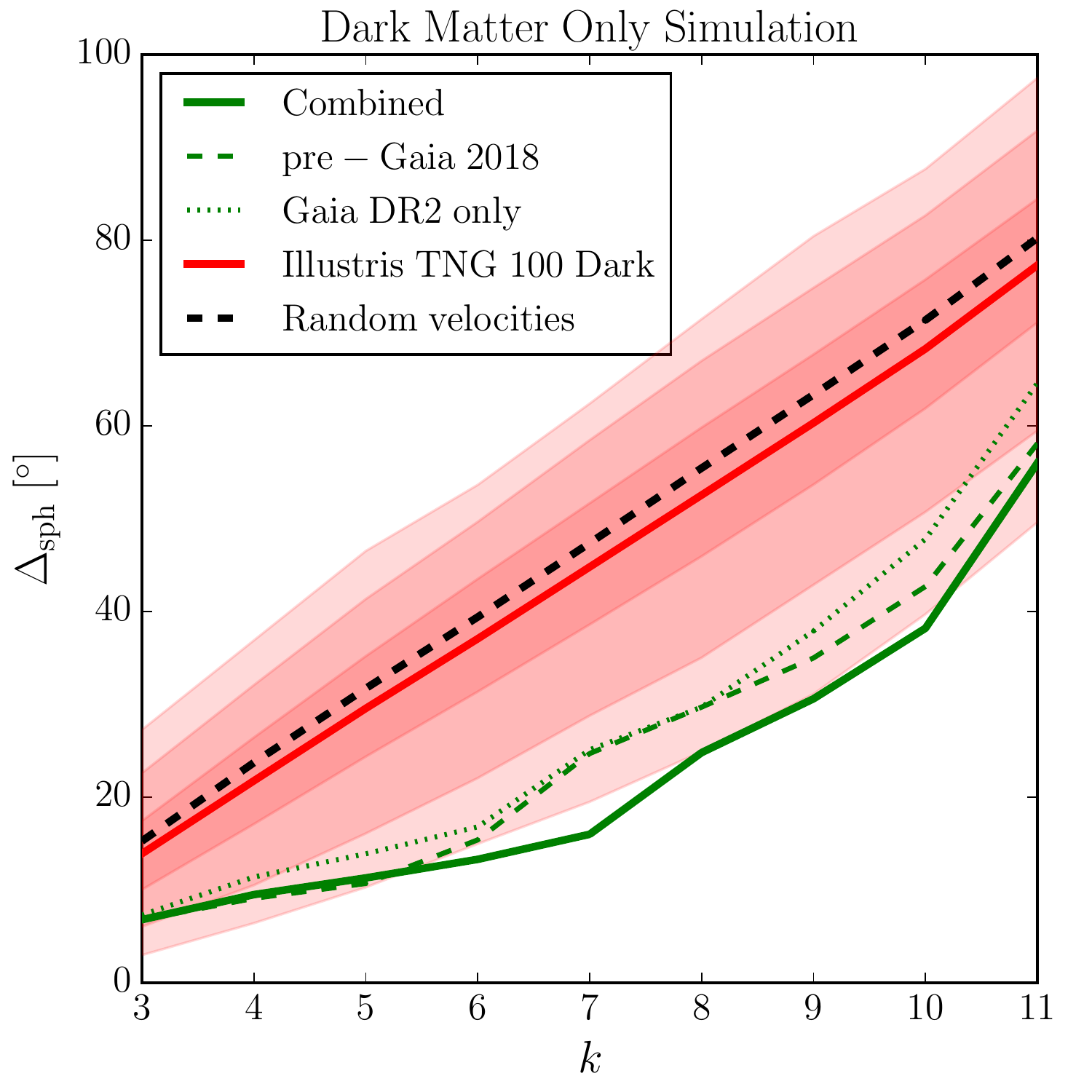}
    \caption{
    Same as Figs. \ref{fig:concen_random} and \ref{fig:concen_best}, but comparing to the distribution of orbital pole directions derived from the hydrodynamical Illustris TNG100-1 cosmological simulation (left panel) and its dark matter only equivalent TNG100-1-Dark (right panel). For comparison, the median $\Delta_\mathrm{std}$\ from the random velocities (black solid line in Fig. \ref{fig:concen_random}) is shown as a black dashed line.
    }
    \label{fig:concen_sim}
\end{figure*}

\begin{figure*}
	\includegraphics[width=\columnwidth]{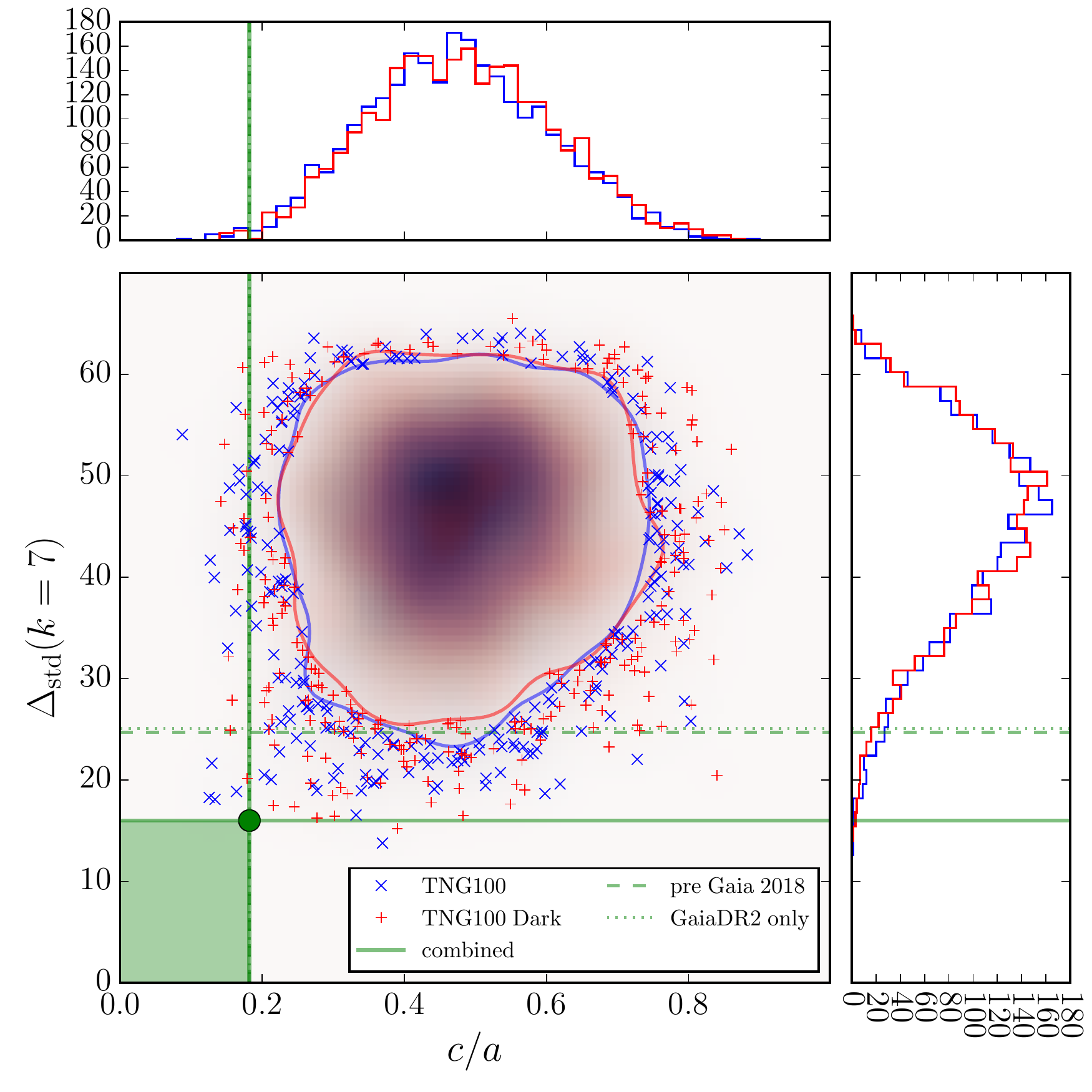}
	\includegraphics[width=\columnwidth]{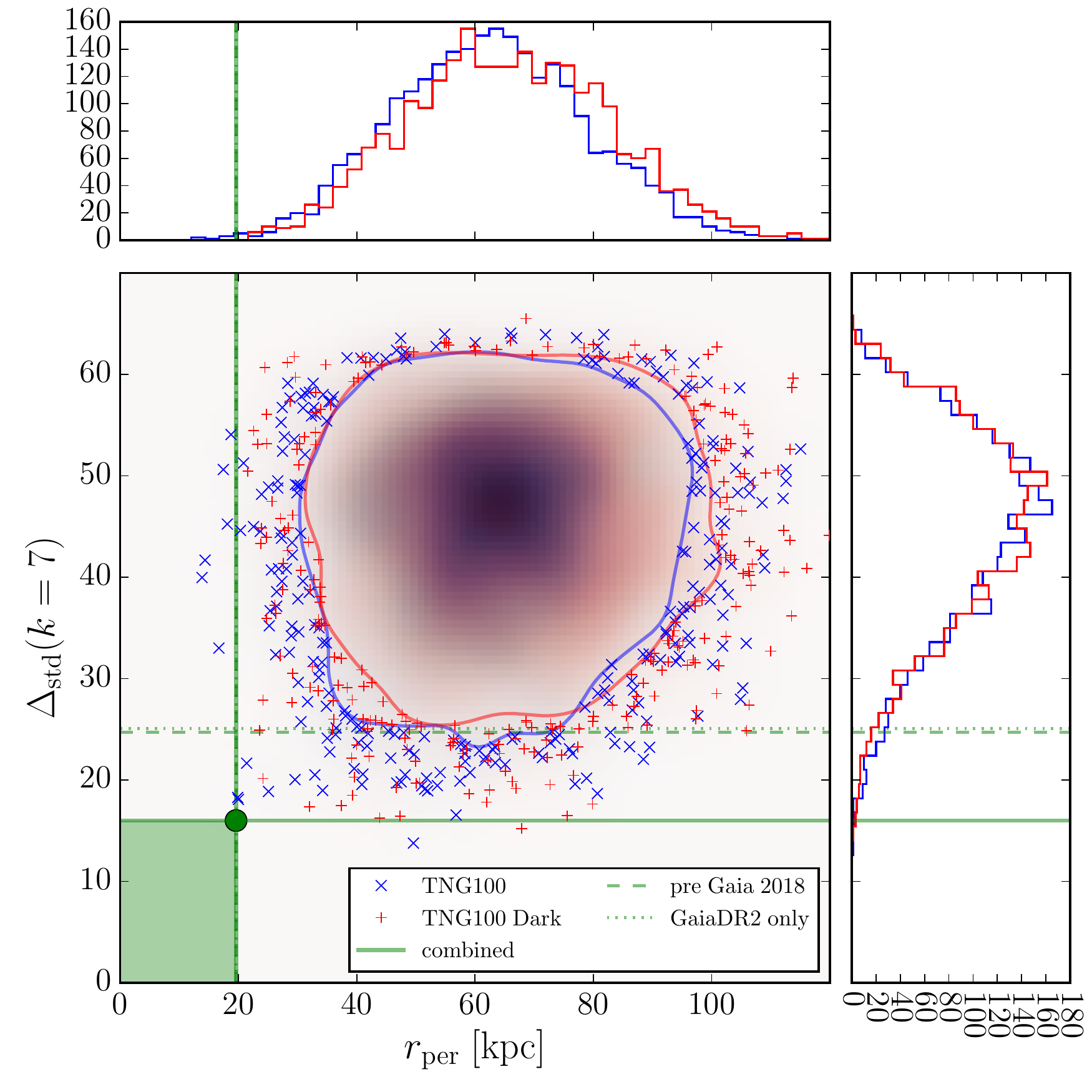}
	\includegraphics[width=\columnwidth]{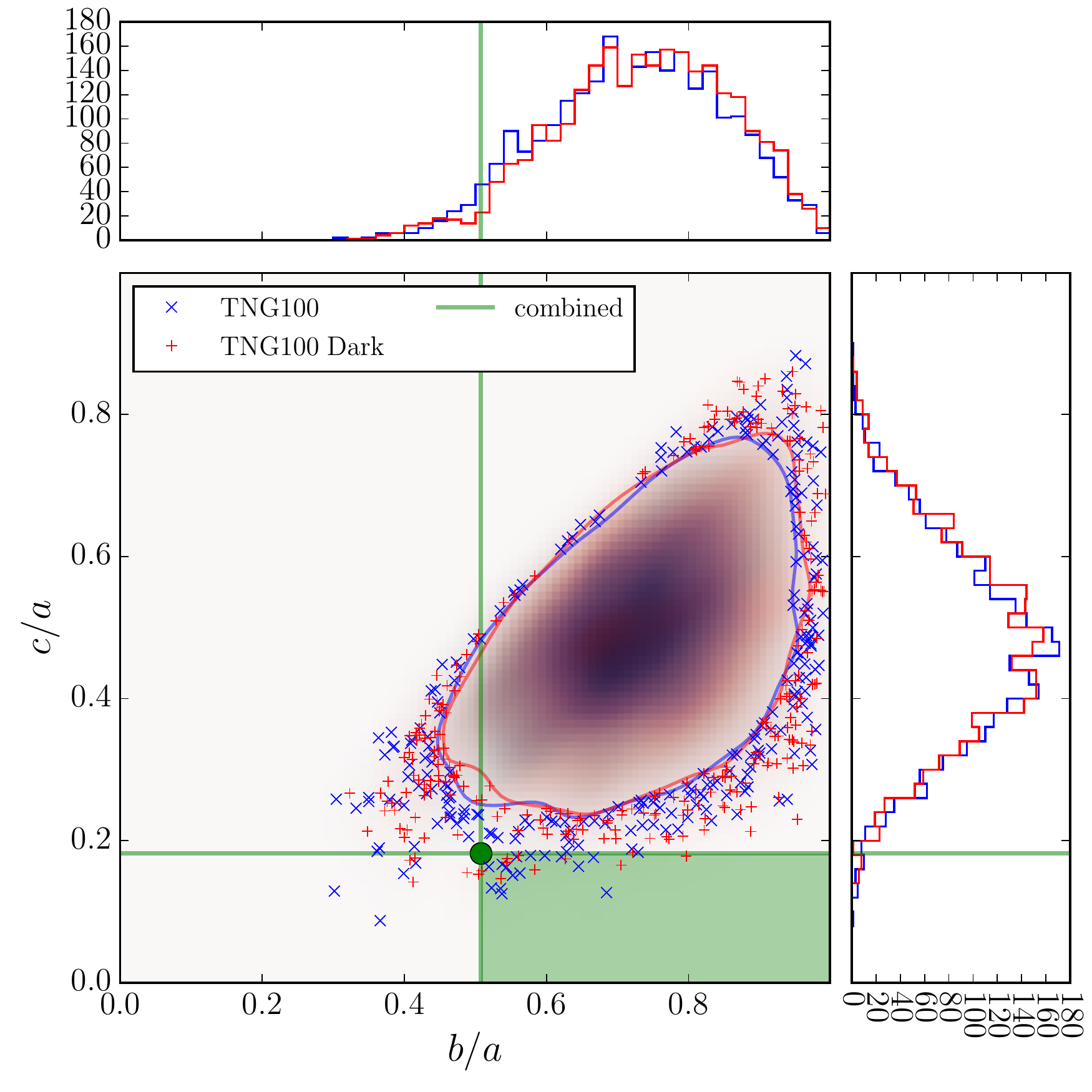}
	\includegraphics[width=\columnwidth]{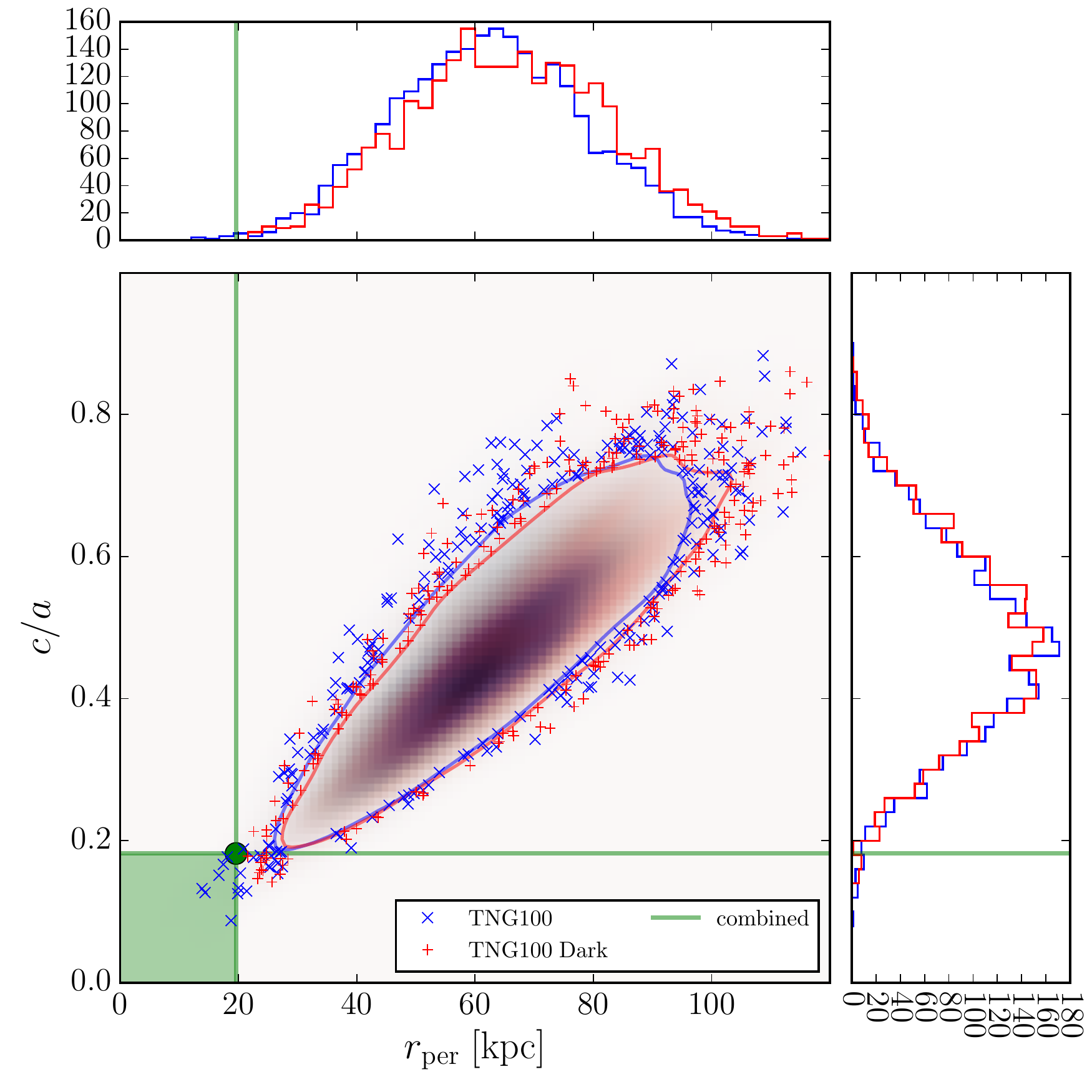}
    \caption{
    Comparison of satellite plane parameters for the observed 11 classical Milky Way satellites (green) with the hydrodynamical cosmological simulation Illustris TNG-100-1 (blue), and its dark-matter-only equivalent Illustris TNG-100-1-Dark (red). The top row plots the orbital pole clustering measure $\Delta_\mathrm{std} (k=7)$ against the relative flattening of the satellite system as measured with the $c/a$ axis ratio (left) and the absolute plane heights $r_\mathrm{per}$\ (right).     The two simulations give very similar results, meaning that the inclusion of baryonic effects does not alleviate the planes of satellite galaxies problem. None of the simulated satellite systems are simultaneously as flattened and as kinematically coherent as the observed classical satellite galaxies of the Milky Way (green shaded area). The lower left panel plots the minor-to-major ($c/a$) vs. the intermediate-to-major ($b/a$) axis ratios, and the lower right panel compares the two measures of flattening $c/a$\ and $r_\mathrm{per}$, indicating their correlation.
    }
    \label{fig:simulations_hydro}
\end{figure*}

The kinematic coherence of the observed satellite galaxies -- as measured with $\Delta_\mathrm{std}$\ for the pre-Gaia 2018, the Gaia-DR2 only, and the Combined samples -- is compared to the results of the hydrodynamical Illustris TNG-100-1 simulation in Fig. \ref{fig:concen_sim}. On average (blue solid line), the simulated systems follow the behaviour of randomly assigned velocities (black dashed line), though with slightly increased coherence. This is in line with previous findings based on different simulations that indicated a close similarity between isotropic and simulated satellite systems \citep{2012MNRAS.424...80P,2014ApJ...789L..24P}. The orbital pole concentrations derived from the pre-Gaia 2018 and the Gaia DR2 only samples lie between the 90 and 99 per cent contours of the simulated systems. Adding the proper motion information provided by Gaia DR2 to previously available high-quality measurements further increases this tension between the observed orbital pole distribution and those expected from cosmological simulations. The Combined sample straddles the 99 per cent contour derived from the simulations, and clearly drops below it for $k= 6$\ and 7. This again strongly suggests a cosmologically unexpected association of at least seven of the 11 classical satellite galaxies with a common plane.

The observed flattening of the system of 11 classical satellites is matched by only 0.75 per cent ($c/a \leq 0.182$) or 0.24 per cent ($r_\mathrm{per} \leq 19.6\,\mathrm{kpc}$) in Illustris TNG100-1. For the dark-matter-only run TNG100-1-Dark, these fractions are even lower, 0.55 and 0.0 per cent, respectively.
These findings are in line with previous studies of other simulations, such as \citet{2019arXiv190402719S} who report that less than one per cent of the \textsc{eagle} simulations have $c/a \leq 0.182$, \citet{2014ApJ...789L..24P} who find 1.0 ($c/a$) and 0.5 ($r_\mathrm{per}$) per cent respectively in the high-resolution, dark-matter-only zoom simulations of the ELVIS project. Slightly higher fractions of 3.4 ($c/a$) and 4.5 ($r_\mathrm{per}$) per cent are found in the lower resolution, dark-matter-only Millennium-II simulation \citep{2014MNRAS.442.2362P}. This might indicate that lower resolution affects the reliability of satellite positions for this kind of analyses.

To resemble the observed satellite plane of the Milky Way, it is not sufficient for a simulated satellite system to reproduce only either the observed kinematic coherence or the observed flattening. Rather, both of these remarkable properties have to be reproduced simultaneously \citep{2014MNRAS.442.2362P,2014ApJ...784L...6I, 2015ApJ...815...19P}. We thus determine the fractions of simulated systems that fulfill both $c/a \leq 0.182$\ and $\Delta_\mathrm{std}(k) \leq \Delta_\mathrm{std}^\mathrm{obs}(k)$\ ($f_{\Delta_\mathrm{std} + c/a}$), as well as both $r_\mathrm{per} \leq 19.6\,\mathrm{kpc}$ and $\Delta_\mathrm{std}(k) \leq \Delta_\mathrm{std}^\mathrm{obs}(k)$\ ($f_{\Delta_\mathrm{std} + r_\mathrm{per}}$). These are listed in Table \ref{tab:deltasph} for the Illustris TNG-100 simulation. Given the already low fractions of systems fulfilling the flattening criteria or the kinematic coherence criteria for $k = [5, ..., 10]$, and the fact that no correlation between flatter and more kinematically coherent orbits is apparent (see Fig. \ref{fig:simulations_hydro}), it is not surprising that the number of simulated systems fulfilling the corresponding flattening and orbital pole concentration criteria simultaneously is negligible. In particular, for $k=7$\ combined orbital poles, \textit{none} of the 2548 simulated systems is simultaneously as flattened as the system of the 11 classical satellite galaxies ($c/a \leq 0.182$ or $r_\mathrm{per} \leq 19.6\,\mathrm{kpc}$) and has as strongly clustered orbital poles as inferred from the combined proper motion sample ($\Delta_\mathrm{std} \leq 16^\circ.0$). This is illustrated in the top panels in Fig. \ref{fig:simulations_hydro}. Analogously, none of the 2564 simulated satellite systems in TNG100-1-Dark fulfills either combination of two criteria.

For other numbers $k$\ of combined orbital poles, the situation is essentially identical. For $k > 3$, at most two simulated satellite systems (0.08 per cent) fulfill the kinematic coherence and at least one of the spatial flattening criteria. The rarity of systems as extreme as the observed plane of satellite galaxies thus does not hinge on a specific choice of $k$\ or which measure of flattening is employed, but rather is a robust finding. Consequently, a look-elswhere effect can not be blamed to be responsible for the found mismatch. 

Combining the new Gaia DR2 proper motions with previous high-quality measurements, and the resulting tighter clustering of orbital poles, thus results in a substantial increase in the degree of tension between the correlated orbital poles of the classical Milky Way satellites and current state of the art cosmological simulations.
At the same time, no significant difference in the frequencies of simulated satellite systems that resemble the observed one are found between the hydrodynamical and the dark-matter-only runs, and neither are there apparent differences in the distributions of the satellite plane parameters as shown in Fig. \ref{fig:simulations_hydro}. This confirms the expectation that the satellite galaxy phase space distribution constitutes a robust test of the cosmological model \citep{ 2005A&A...431..517K,2018MPLA...3330004P}, indicates that previous studies using dark-matter-only simulations to compare to the VPOS remain valid, and confirms earlier indications that the modelling of baryonic processes in cosmological simulations does not alleviate the planes of satellite galaxies problem \citep{2017MNRAS.466.3119A,2015ApJ...815...19P,2017AN....338..854P,2019ApJ...875..105P, 2018MNRAS.476.1796S}.

\subsection{Different measures of orbital pole clustering}

\begin{table}
	\centering
	\caption{The radii of circles on the sphere that contain the most concentrated $k$\ orbital poles for the Combined sample, either using vectors (orbital poles) or axes (orbital axis, i.e. ignoring direction of orbit).
	}
	\label{tab:vecaxial}
	\begin{tabular}{lcc}
  \hline
$k$ & $\Delta_\mathrm{vec}\ [^\circ]$ & $\Delta_\mathrm{axial}\ [^\circ]$ \\
  \hline
3   & 8.1 & 8.1 \\
4   & 11.6 & 10.9 \\
5   & 13.8 & 13.8 \\
6   & 17.8 & 15.6 \\
7   & 19.9 & 17.7 \\
8   & 41.7 & 19.9 \\
9   & 44.9 & 41.3 \\
10 & 56.1 & 44.9 \\
11 & 90.4 & 56.1 \\
  \hline
	\end{tabular}
\end{table}

\begin{figure*}
	\includegraphics[width=55mm]{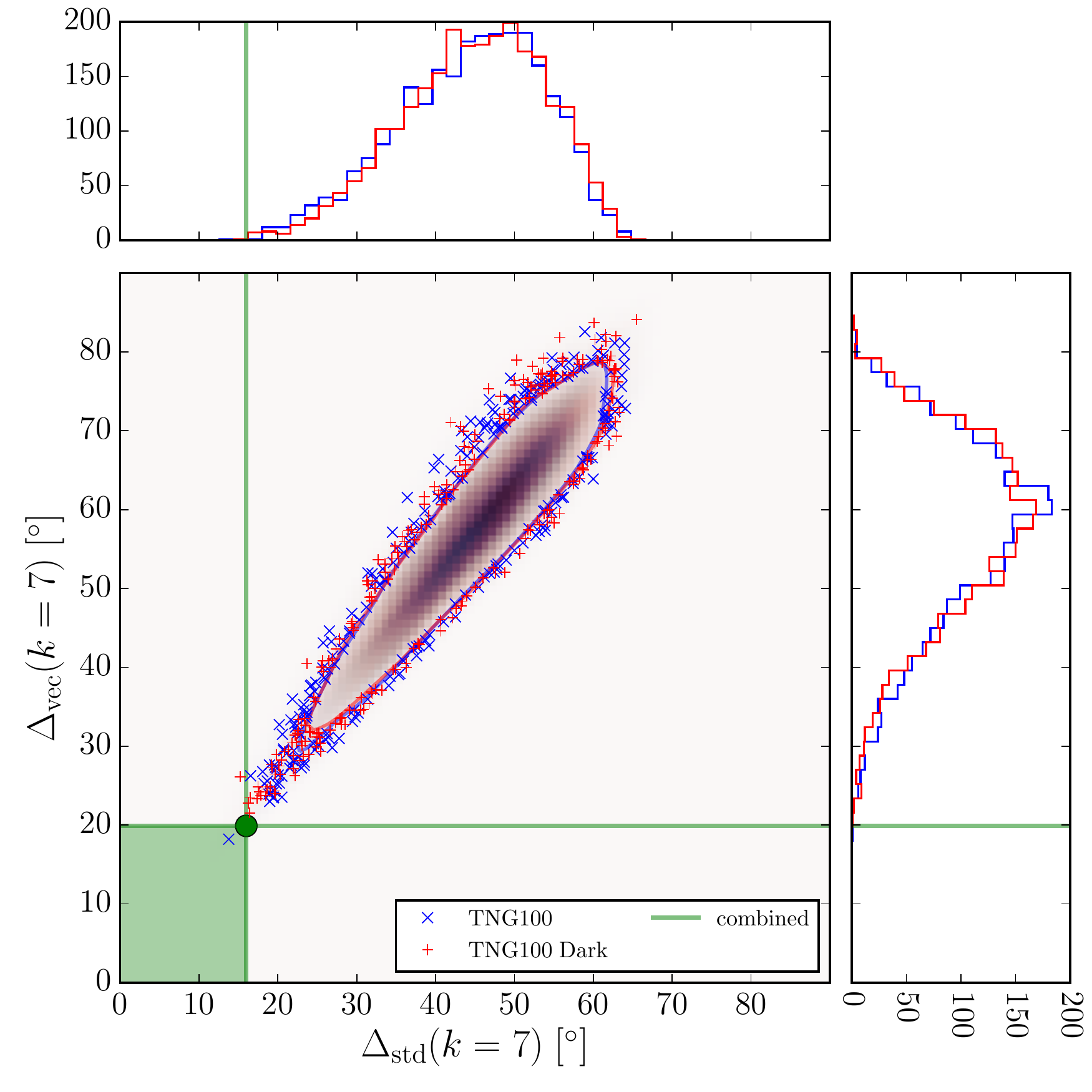}
	\includegraphics[width=55mm]{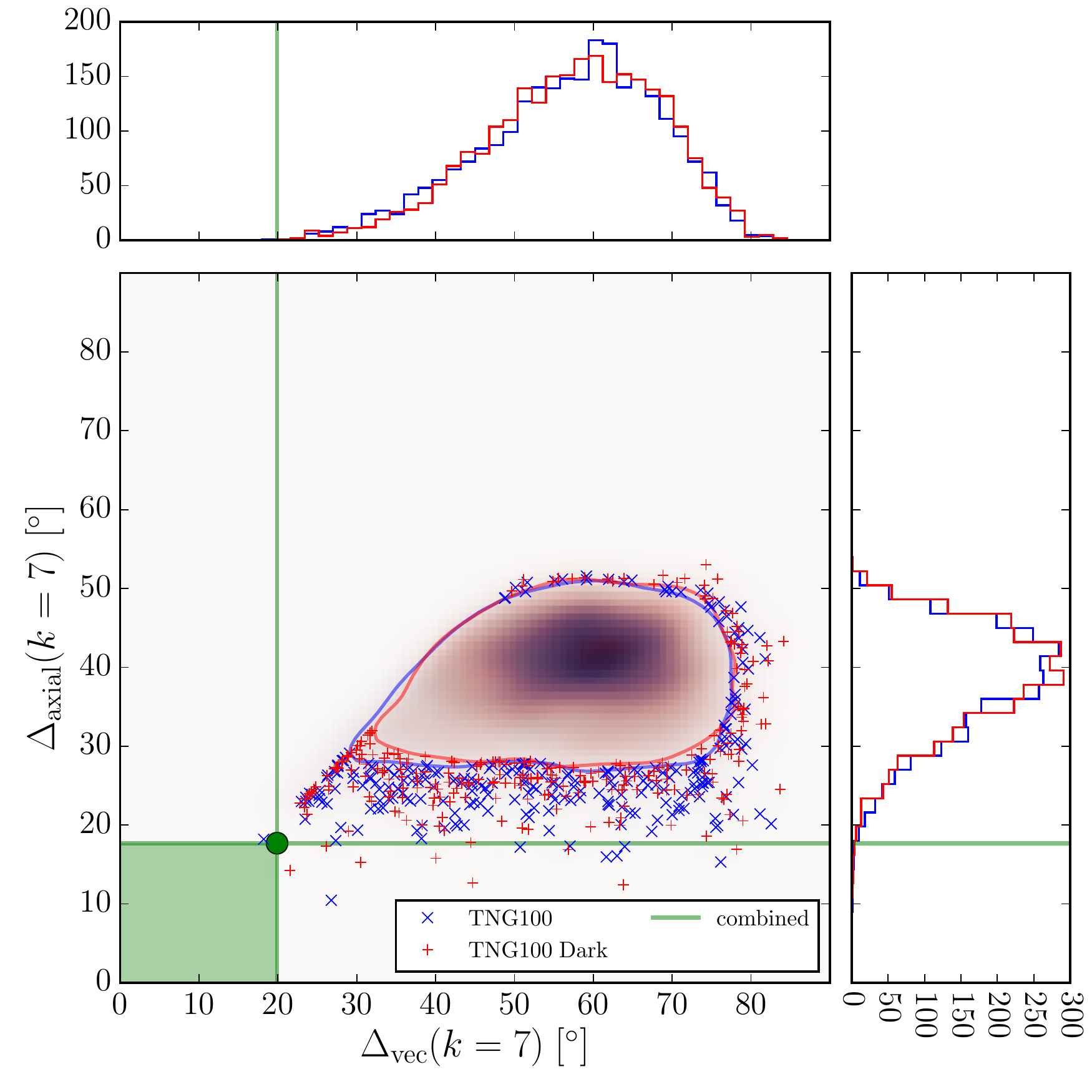}
	\includegraphics[width=55mm]{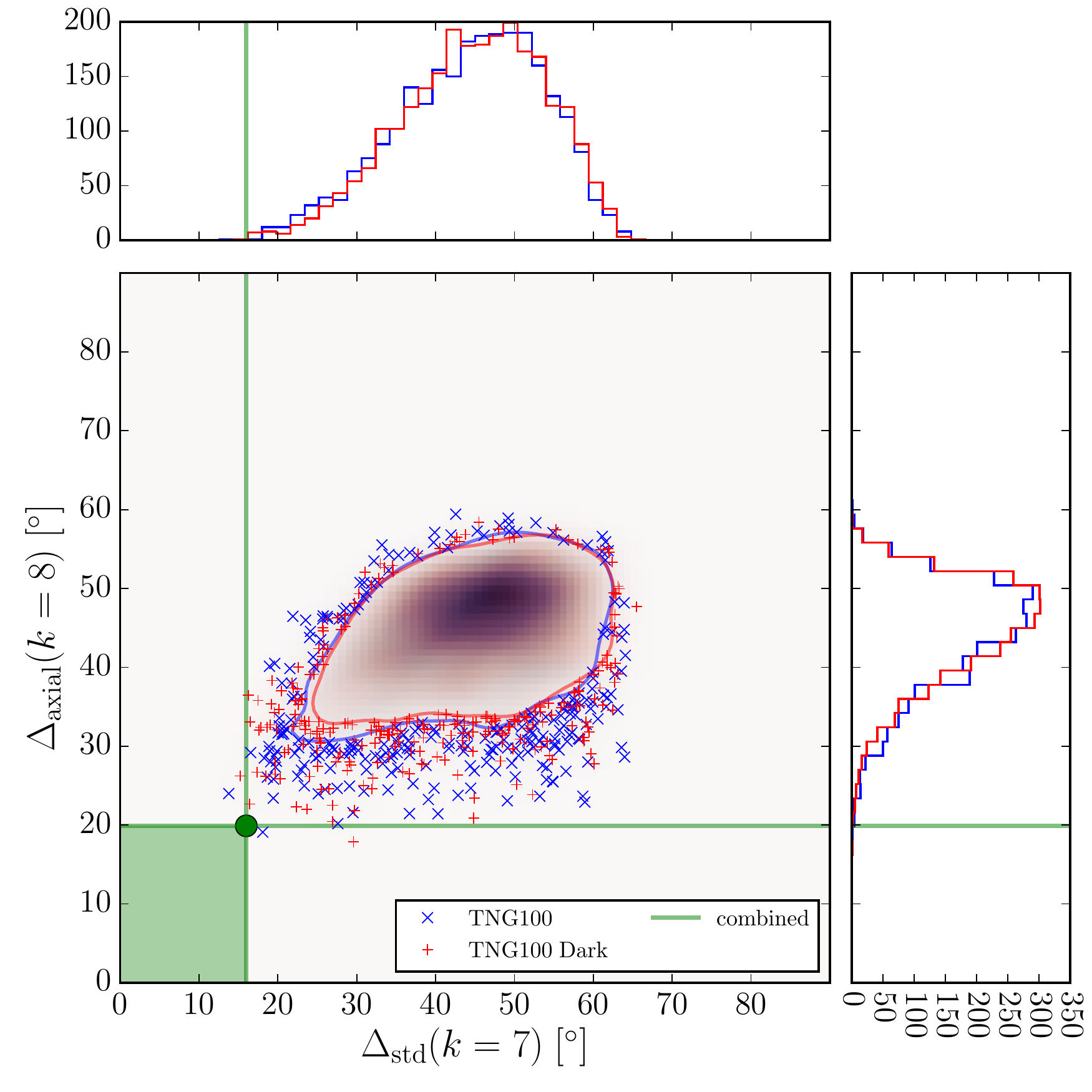}
    \caption{
Comparison of three measures of orbital pole clustering in the cosmological simulations, plots analogous to those in Fig. \ref{fig:simulations_hydro}. 
The left panel compares the commonly used spherical standard distance $\Delta_\mathrm{std}(k)$\ of the $k$\ most concentrated poles with the radius $\Delta_\mathrm{vec}(k)$\ of the smallest circle containing $k$\ orbital poles and shows them to be highly correlated (here shown for $k=7$). The middel panel compares $\Delta_\mathrm{vec}(k)$\ with $\Delta_\mathrm{axial}(k)$\ which considers only orbital axes (i.e. ignores the sign of the angular momentum) when finding the smallest circle containing $k$. No strong correlation is present, except in cases where no counter-orbiting satellites are found within the direction opposite of the originally identified circle resulting in the limiting line $\Delta_\mathrm{vec}(k)$ = $\Delta_\mathrm{axial}(k)$. The right panel compares $\Delta_\mathrm{std}(k)$\ with $\Delta_\mathrm{axial}(k)$ (for $k=8$) and reveals a similar trend, though it also shows that for the highly correlated orbital poles of the Milky Way satellites (green lines), either measure finds simulated satellite systems with at least as well aligned orbits as the observed classical satellites of the Milky Way to be exceedingly rare.
    }
    \label{fig:compare_std_vec_axial}
\end{figure*}

Recently, \citet{2019arXiv190402719S} introduced another measure of orbital pole clustering. For a given distribution of orbital poles, it finds the minimum opening angle $\Delta_\mathrm{axial}(k)$ of a cone that contains $k$\ orbital axes (the axis parallel to the orbital poles). The method thus does not discriminate between co- and counter-orbiting satellites and instead considers all satellites orbiting along a common plane to be aligned. 
Here, we investigate how this measure compares to the more standard spherical standard distance $\Delta_\mathrm{std}$, as well as with an analogous measure that {\it does} consider the direction of the orbital poles, i.e. $\Delta_\mathrm{vec}$. In particular, we want to investigate whether the choice of measure can bias the frequency of finding analogue systems in cosmological simulations.

Our algorithm starts with a near-uniform distribution of 10000 points on a sphere. To find $\Delta_\mathrm{vec}$, it measures the angular seperation $\alpha_{i,j}$\ between each point $i$\ and every satellite galaxy $j$. For every point the 11 angles are ranked, and for any number $k = [3, ..., 11]$\ the smallest such angle among the 10000 points is determined. This is the opening angle of the smallest cone containing $k$\ of the 11 orbital pole directions. To determine $\Delta_\mathrm{axial}$, the algorithm instead measures the angular seperations from the orbital axes, i.e. $\alpha'_{i,j} = \min(\alpha_{i,j}, 180^\circ-\alpha_{i,j})$. While strictly speaking this is only an upper limit on the actual minimum circle due to the finite separation between the 10000 points, it provides sufficient accuracy for our purposes, in particular considering that the individual orbital pole directions of the observed Milky Way satellite galaxies are still uncertain to $\gtrsim 1^\circ$\ (see $\Delta_\mathrm{pole}$\ in Table \ref{tab:individualpoles}). Furthermore, this procedure as well as the chosen number of uniformly distributed points is identical to that used in \citet{2019arXiv190402719S}, which thus makes our results immediately comparable to theirs.

Table \ref{tab:vecaxial} compiles the results of applying these measures to the distribution of orbital poles for the observed Milky Way satellite galaxies from Table \ref{tab:individualpoles}, based on the Combined sample of proper motions. In practice, for the observed classical satellite galaxies of the Milky Way, $\Delta_\mathrm{vec}$\ and $\Delta_\mathrm{axial}$\ differ mostly due to the flipping of the orbital pole of Sculptor, which is counter-orbiting relative to the main cluster of poles, such that $\Delta_\mathrm{vec} (k) \approx \Delta_\mathrm{axial} (k+1)$\ for $k > 5$.

These observed values can then be compared to those of simulated satellite systems in the Illustris TNG-100 and TNG100-Dark simulations. This is done in Fig. \ref{fig:compare_std_vec_axial}. The left panel in that figure compares the commonly used spherical standard distance $\Delta_\mathrm{std}(k)$\ of the $k$\ most concentrated poles and the radius $\Delta_\mathrm{vec}(k)$\ of the smallest circle containing $k$\ orbital poles (shown for $k=7$). These two measures are clearly highly correlated, and also identify the same one simulated systems in the Illustris TNG100 simulation as being as coherent as the observed classical satellite galaxies of the Milky Way. In contrast, the middle panel reveals that measuring the clustering of orbital axes (i.e. ignoring the sign of the angular momentum) via $\Delta_\mathrm{axial}(k)$\ does not show a strong correlation with $\Delta_\mathrm{vec}(k)$. Only in those cases where no counter-orbiting satellites are found within the direction opposite of the originally identified circle $\Delta_\mathrm{vec}(k)$ = $\Delta_\mathrm{axial}(k)$, otherwise $\Delta_\mathrm{axial}(k)$\ results in smaller values and there is considerable scatter for a given $\Delta_\mathrm{vec}(k)$. The axial measure is also less constraining as it identifies more simulated systems as more extreme than the observed one (below the horiziontal green line corresponding to the value for the observed Milky Way system) in terms of the corresponding vectorial measure (vertical green line). This difference means that the systems identified via $\Delta_\mathrm{axial}$\ do not actually resemble the observed situation since they do not have a majority of seven satellites co-orbiting in the same sense.
Finally, as becomes apparent from the right panel of Fig. \ref{fig:compare_std_vec_axial} the situation is similar when comparing $\Delta_\mathrm{std}(k=7)$\ with $\Delta_\mathrm{axial}(k=8)$. However, for our specific situation either of the measures of orbital coherence give similar results: simulated satellite systems with at least as well aligned orbits as the observed classical satellites of the Milky Way are exceedingly rare (at most one or two out of $\approx 2500$\ simulated systems).

\section{Conclusions}

The Milky Way and several other major galaxies are surrounded by planes of satellite galaxies, flattened distributions of satellite dwarf galaxies with correlated kinematics. For external systems, line-of-sight velocities of approximately edge-on satellite planes are indicative of common orbital directions of satellites along the planes, but tangential motions are unconstrained. The Milky Way, in contrast, offers the unique advantage that proper motion measurements provide access to full three-dimensional velocities, thus enabling a study of the degree of orbital coherence of satellite galaxies with the plane. Combining the best-available proper motion measurements from a variety of ground- and space-based sources has revealed a preference for the 11 brightest, ``classical'' satellite galaxies to co-orbit along the plane defined by their current positions \citep{2013MNRAS.435.2116P}.

With Gaia DR2, a new, entirely independent and homogenous set of proper motions for the 11 classical satellite galaxies has become available \citep{2018A&A...616A..12G}. We have used this data set, in combination with previous measurements, to test previous claims and quantify the effect of including the Gaia DR2 proper motions on the amount of orbital correlation of the 11 classical satellite galaxies and its degree of tension with cosmological expectations.

One of our main results is that the Gaia DR2 proper motions confirm previous findings. The distribution of orbital poles of seven of the 11 classical satellite galaxies is strongly clustered in the direction of the normal vector to the VPOS, as expected for a rotating structure. In addition to the six satellites previously identified to co-orbit -- LMC, SMC, Draco, Ursa Minor, Fornax, and Leo\,II -- the new Gaia DR2 now also places the Carina dwarf spheroidal satellite galaxy firmly into this group. As previously known, Sculptor also orbits along the same plane but in the opposite direction, such that a total of eight out of the 11 brightest Milky Way satellites have orbital planes aligned to within $20^\circ$\ of a common direction. With an angle of $\theta_\mathrm{VPOSclass} = 22^\circ.1$, the average direction of the sample of $k = 7$\ most-concentrated orbital poles is also the one that aligns best with the VPOSclass normal vector, and aligns even closer with the VPOSall normal vector direction. Such a close alignment between the short axis of the spatial distributon of the satellite system and the average orbital axis of the co-orbiting satellites is not expected if the velocities were random, but is in line with the satellite galaxies indeed preferentially orbiting along the common plane inscribed by their current positions.

The combination of the best-available proper motions from Gaia DR2 and other sources thus results in the most tightly clustered distribution of orbital poles measured to date, with a significance of the clustering to over 99.9 per cent compared to velocity directions drawn from isotropy. The degree of orbital pole concentration has continuously increased with increasing precision of proper motion measurements, as expected if the underlying distribution contains a true physical correlation. Such a strong underlying correlation of the orbital planes of a majority of the 11 classical satellite galaxies is also consistent with the clustering of their three to seven best-aligned orbital poles, which show spherical standard distances that trace those expected from assuming that the underlying orbital planes are as well aligned as possible with the VPOS given the current positions of the satellites.

Intriguingly, most of the satellites sharing a common orbital plane to within $\approx\,20^{\circ}$\ also have similar specific angular momenta, within 10 to 30 per cent of the value predicted by \citet{1995MNRAS.275..429L} based on assuming that they originate from a common orbit. Leo\,II and Fornax are the two exceptions with almost twice the average specific angular momentum of the other correlated satellites. This, as well as its three times larger distance from the Milky Way, thus makes it questionable whether Fornax could plausibly be considered as a past satellite galaxy of the LMC, as claimed by some recent studies \citep{2019arXiv190401028P,2019arXiv190702979J}. As we have shown, many of the 11 classical satellites have orbital poles that are closely aligned with that of the LMC and also fall along a common great circle on the sky. However, some of these (e.g. Draco, Ursa Minor) are on the opposite side of the Milky Way and thus can not be considered as past LMC satellites in a first infall scenario. We therefore caution against basing confirmation of a past association as a satellite of the LMC on merely the alignment of their orbital poles and their closeness to the LMC in projection on the sky. Rather, careful energy and angular momentum considerations in the spirit of \citet{1995MNRAS.275..429L}, as well as comparisons to $N$-body models \citep[e.g.][]{2011ApJ...742..110N,2017MNRAS.465.1879S,2019arXiv190709484E} are required to determine whether a common infall is possible and plausible.

Using the new information provided by the Gaia DR2 proper motions, we also re-assess the degree of tension between the phase-space distribution of the 11 classical satellite galaxies and cosmological expectations based on the $\Lambda$CDM model. For this, we compare with the hydrodynamical Illustris TNG100-1 simulation as well as its dark-matter-only counterpart, with identical results. Adding the Gaia DR2 data to previous proper motion information reduces the frequency of simulated systems with the same degree of orbital pole correlation from 2-3 per cent to under 0.1 per cent for the seven most-correlated satellites. When combined with the requirement that the simulated satellite system is simultaneously as flattened as the observed one, the frequency of systems as extreme as the observed one essentially drops to zero. The Gaia DR2 proper motions thus do not alleviate, but rather compound the Plane of Satellite Galaxies Problem of the Milky Way. The fact that three of the closest massive host galaxies with well studied satellite galaxy systems -- the Milky Way, M31, and Centaurus A -- all have planes of satellites suggests not only that these structures are more the rule than the exception, but also that the $\Lambda$CDM cosmological model may be significantly threatened by this very situation.

\section*{Acknowledgements}
This work has made use of data from the European Space Agency (ESA) mission {\it Gaia} (\url{https://www.cosmos.esa.int/gaia}), processed by the {\it Gaia} Data Processing and Analysis Consortium (DPAC,
\url{https://www.cosmos.esa.int/web/gaia/dpac/consortium}). Funding for the DPAC has been provided by national institutions, in particular the institutions participating in the {\it Gaia} Multilateral Agreement.
This research has made use of NASA's Astrophysics Data System, of the IPython package \citep{PER-GRA:2007}, SciPy \citep{jones_scipy_2001}, NumPy \citep{van2011numpy}, matplotlib, a Python library for publication quality graphics \citep{Hunter:2007}, and Astropy, a community-developed core Python package for Astronomy \citep{2013A&A...558A..33A}. The acknowledgements were compiled using the Astronomy Acknowledgement Generator.

%%%%%%%%%%%%%%%%%%%%%%%%%%%%%%%%%%%%%%%%%%%%%%%%%%

%%%%%%%%%%%%%%%%%%%% REFERENCES %%%%%%%%%%%%%%%%%%

\bibliographystyle{mnras}
\bibliography{VPOS_GaiaDR2_filltex}

%%%%%%%%%%%%%%%%%%%%%%%%%%%%%%%%%%%%%%%%%%%%%%%%%%

% Don't change these lines
\bsp	% typesetting comment
\label{lastpage}
\end{document}